\documentclass[prb,amssymb,twocolumn]{revtex4}
\usepackage{myDEF}

\begin{document}
\title{Coherent Multiple Backscattering of Waves in Random Medium\\ beyond the Diffusion Approximation}
\author{Peter Slavov\protect\footnote{\ petersl@phys.uni-sofia.bg}\\
{\it University of Sofia, Faculty of Physics}}
\date{\today}
\begin{abstract}
In this paper the validity of the \DA\ for multiple \sc\ of classical waves
in random medium in different regimes is investigated, with emphasize to \WL\
effects. Many principle topics are discussed once again. We obtain new \ex s
for the \ic\ and \coh\ \MSC\ intensities for the more general case of large
\sc\ angles and various observation points inside the medium with moving
scatterers and absorption. (The results, we reached, assuming white-noise
disorder for the \IC, contain one-dimensional integration in its final
analytical form). Also new self-consistent formulas for \ic\ and \coh\
intensities for fixed multiplicity orders of \sc\ are given and their
lineshapes are graphically represented. Through these formulas we clearly
demonstrate that the impurities motion reduces the contribution to the \I\
of the \sc s with large multiplicity orders.
The total \sc\ intensities in $4\pi$ solid angle are calculated for the
general case and compared with their \DA. For \coh\ intensities it is shown
that a drastic deviation from the exact results occurs. It turns out that the
region of angles, for which the observed lineshape of the \WL\ effect tends to
the \ic\ background, give the main contribution to the total \coh\ scattered \I.
The last quantity happens to depend linearly instead of predicted quadratic
dependence on the perturbation parameter, $1/k\el$.
\end{abstract}

\maketitle

\section{Introduction.}
Anderson \loc\ \cite{A1,M1,M2,W,VW,AALR,LR} was predicted and demonstrated as a \loc\ of the conducting electrons. The later understanding that it is not only a quantum effect but applies as well to classical (sound or electromagnetic) waves \cite{Bar,SJ1,SJ2,Gb,A2,TzIsh} stimulated a new interest in the study of propagation and \MSC\ of waves in \RM\ \cite{St1,StC,AWM,St,St2,McKSJ,McKSJ2,PSh}.

The discussion of wave \loc\ in disordered media usually is  concentrated into two characteristic regions:

(1)\ \underline{Region of strong \loc.} This is the region, where the Ioffe-Regel criterion $\el\!\simeq\!\lambda$ is satisfied. Here $\lambda$ is the {\it wavelength} of scattered wave and $\el$ is the \EMFP. In this case the diffusion of the wave in the medium should tend to zero when the disorder becomes larger than a critical value. The transition to strong Anderson \loc\ in both cases of quantum particles and classical waves is caused by interference effects which reduce the value of the diffusion constant until it vanishes when $\lambda$ becomes of the order of $\ell$. However this regime is poorly known in the interesting for us case of
\MSC\ of classical waves.

(2)\ \underline{Region of \WL.} In this case $\el\!\gg\!\lambda$ and the phenomenon of \WL\ is identified with the effect of multiple \CBSC\ of sound or light by the disordered medium. Weak \loc\ of light was studied extensively theoretically \cite{Gb,TzIsh,St1,StC,AWM,St,St2,McKSJ,McKSJ2} and experimentally \cite{AL,WM,KREF}. The main result is that in a medium with a random distribution of  static scatterers there is a peak in the \I\ of the multiple scattered wave in a narrow solid angle around the backward direction. This peak of light \I\ is almost twice as high as the \ic\ background \I. The increase of \I\ in backward direction is due to \coh\ interference effects.

All analytical results for the \CBSC\ \I\ were obtained in \DA. Most of the authors discussing the \CBSC\ of  light in \RM\ clearly understood that this \app\ is justified only at small angles, but at large angles (beyond the \BSC\ cone) it ceases to be valid \cite{Gb,AWM,McKSJ}. Therefore a careful investigation of the validity of the \DA\ in different regimes is needed. Similarly it is desirable to have \ex s of the \coh\ and \ICSC\ intensities without any restriction to small angles around the backward direction. In the present paper we realize these two goals considering the explicit example of semi-infinite homogeneous medium with white noise disorder of point scatterers.

We started this work, when trying to find answers to some questions concerning the behaviour of the \MSC\ \I\ in depth of the \HM:

Is there any nonzero asymptotic value of the \CBSC\ \I\ at infinite depth into the \hom\ half-space? What happens with the typical lineshape of the \CBSC\ peak at various depths in the medium?

The main point in our further reasoning is the understanding that the wave \loc\ is a phenomenon, which take place inside the \sc\ medium. It is clear that, if the observation point is deeper localized into the medium, the \MSC\ \I\ becomes more isotropic with respect to the \sc\ angles. From this point of view, the investigation of the angular dependence of both \coh\ and \ICSC\ intensities for $4\pi$-geometry comes out to be an important problem. Keeping in mind these considerations, we start with obtaining the \ex\ for the \FI\ of the \sc\ \pr, based on the well known \DA. Further we find the \MSC\ \I\ in this approach. The result is an increase of the \ic\ \I\, at infinity, to an asymptotic value, twice higher than that one on the surface at backward direction, and that the \coh\ \I\ tends to zero at infinite depth inside of the medium. On the other hand, as we mentioned above, the \DA\ brings a significant error, when the angular dependence of \coh\ \I\ is extended to $4\pi$-geometry. We find that an additional term has been lost in the \I\ \ex, when the \DA\ has been used. In fact, this term appears, when we proceed with the exact(not approximated) \lpr. Using the exact \pr, we reach new results, which demonstrate that the \coh\ \I\ at infinite depth inside of the medium tends to zero again, and the nondiffusion term turns out to be several times greater than the diffusion one. It must be noted here, that our calculations of scattered \I\ were carried out, after imposing the correct image boundary conditions on the \sc\ \pr\ \cite{PS,Mark}.

Further we investigate the contribution to the \sc\ \I\ of the \ic\ and \coh\ \MSC\ of a given multiplicity order. A comparison is made of the exact results and the \DA\ in this way we explicitly demonstrate that the motion of the impurities reduces the contribution of the \coh\ \sc\ with large number of \sc s. New formulas are obtained also for the total \sc\ intensities.

\section{Basic notations.}
\subsection{The perturbation theory.}
We shall begin with a short review of the \MSC\ of light and the commonly accepted \app s. The basic instruments, for description of propagation of waves in \RM\ with small fluctuations, are adopted quantum field theory methods such as the \GF\ technique and diagram approach.
The above mentioned methods applied to \MSC\ of waves and are known as {\it impurity-averaged perturbation theory in the small parameter $1/k\el$}\ and they are the ground for a quantitative theory of the \MSC\ of light. According to this theory there are two general \ieq s, which one faces with -- Dyson's and \BSEQ s.

In what follows, we shall neglect the vector structure of the electromagnetic field (polarization effects) and we shall consider only scalar monochromatic waves. For electromagnetic waves, the randomness is caused by fluctuations of the dielectric constant, so that the disorder in the system will be described by a random part of the relative dielectric constant $\epsilon(\rt)\!=\!1\+\ept{}$. The \ex\ for \RGF\ can be directly derived from the scalar \WE:
\beq
{\bf\square} u(\rt)-\ept{}\dtdt{}{u(\rt)}
={\rm j}(\rt)\;\;\;\;;\;\;\;\;\;\;\;\rt:=(\rs,\ttt) .
\eeq
We shall suppose that $\ept{}$ is a complex variable $\ept{}=\epr{}+i\ep''$ such that its real part is a Gaussian random field $\av{\epr{}}_{ensemble}\!=\!0$ and the imaginary part is a constant due to absorption processes -- $\av{\ep''}_{ensemble}\!=\!\ep''$.

The type of disorder of the medium is given by {\it the \CF\ of the scatterers (\IC)}:
$$\PErt{}0:=\av{\epr{}\epr0}$$

The random \sc\ process is described by \RGF\ $\Gr{}0$, which satisfies the \eq:
\beq[w1]
\Br{{\bf\square}_{\rt}-i\ep''\dtdt{}{}}\Gr{}0-\epr{}\dtdt{}{\Gr{}0}=
\DE(\rt\-\rt_0) .
\eeq
The unperturbed (without the fluctuating part $\epr{}$) \WE\ is:
\beq[w2]
\Br{{\bf\square}_{\rt}-i\ep''\dtdt{}{}}\Go{}0=
\DE(\rt\-\rt_0) .
\eeq
The \GF\ technique translates the problem for obtaining the \RGF\ from the language of differential \eq s to the language of \ieq s. Moreover, if we use this technique, the inclusion of source distribution becomes a trivial operation. Further we represent the \RGF\ as a sum of the unperturbed \GF\ $\Go{}0$ and a perturbation term. The last one can be generated, if the fluctuating term in \Eq{w1} is considered as a source distribution for \Eq{w2}. Thus the \RGF\ is:
\beq[per]
\Gr{}0=\Go{}0+\int\!\Go{}1\epr1\dtdt1{\Gr10} \drt_1
\eeq
When performing the average over the disorder $\epr{}$, we have to take into account that for a Gaussian fluctuation all odd statistical moments are zero and all even statistical moments are products of the \IC s:
\WBeq[av]
\lefteqn{\Gm{}0=\Go{}0+k_0^4\int\!\!\Go{}1\PErt12\Go12\Go20\drt_1\drt_2+}\NN
& &+k_0^8\!\int\!\!\Go{}1\PErt12\Go12\Go23\PErt34\Go34\Go40
\drt_1\drt_2\drt_3\drt_4+ \NN
& &+k_0^8\!\int\!\!\Go{}1\Go12\PErt13\Go23\Go34\PErt24\Go40
\drt_1\drt_2\drt_3\drt_4+ \NN
& &+k_0^8\!\int\!\!\Go{}1\PErt14\Go12\PErt23\Go23\Go34\Go40
\drt_1\drt_2\drt_3\drt_4+ \NN
& &+k_0^{12}\!\int\dots\,,
\WEeq
where $k_0$ is the magnitude of the wave vector of the incident light. In this expansion of the \AGF, each term corresponds to some order of the perturbation or equivalently to some diagram:

%
\def\Hline#1#2#3{
\put(#1,0){%
\put(0,0){\line(1,0){#2}}\multiput(-0.6,0)(#3,0)2\BOX
\put(0,4){\mbrt{}}\put(#3,4){\mbrt{_0}} }%
}
\def\COR#1#2#3{%
\put(#1,0){\put(8,0)\ArC\put(1.5,-3){\mbrt{_{#2}}}
\put(14.5,-3){\mbrt{_{#3}}}} }%
\unitlength=0.8mm \lth{0.4pt}
\bfigW
\begin{picture}(218,20)(1,-8)
\put(0,0){{\lth{2pt}\line(1,0){10}}}\multiput(-0.6,0)(11,0)2\BOX
\put(0,4){\mbrt{}}
\put(11,4){\mbrt{_0}}
\put(16,0)\EQUAL
\Hline{22}{10}{11}
\multiput(37,0)(33,0)2\PLUS
\Hline{42}{23}{24}
\COR{45.5}12
\Hline{75}{46}{47}
\COR{79}12
\COR{101}34
\multiput(126,0)(41,0)2\PLUS
\Hline{131}{31}{32}
\COR{135}13
\COR{142}24
\put(188,0){
\Hline{-16}{32}{33}
\COR{-8}23
\put(0,6.93)\PT
\multiput(-1.21,6.88)(2.42,0)2\PT
\multiput(-2.41,6.72)(4.82,0)2\PT
\multiput(-3.59,6.46)(7.18,0)2\PT
\multiput(-4.74,6.09)(9.48,0)2\PT
\multiput(-5.86,5.63)(11.72,0)2\PT
\multiput(-6.93,5.07)(13.86,0)2\PT
\multiput(-7.95,4.42)(15.90,0)2\PT
\multiput(-8.91,3.69)(17.82,0)2\PT
\multiput(-9.80,2.87)(19.60,0)2\PT
\multiput(-10.61,1.98)(21.22,0)2\PT
\multiput(-11.35,1.02)(22.70,0)2\PT
\multiput(-12,0)(24,0)2{\PTL}
\put(-12,-3){\mbrt{_1}}\put(12,-3){\mbrt{_4}}
\put(21,0)\PLUS
\put(29,0)\DOTS
}
\end{picture}
\caption{\mbox{Diagrammatic representation of the \AGF.}}
\label{psum}
\efigW 

It is convenient to classify the different \dg\ appearing on (Fig.~\ref{psum}) into reducible and irreducible \dg\ \cite{RKT}. This way of sorting all the \dg, enables us to rewrite the infinite sum \Eq{av} in a compact form -- Dyson's \ieq:
\Beq[DEQ]
\lefteqn{\Gm{}0=\Go{}0+}\NN
&&+\int\!\Go{}1\DSrt12\Gm20\drt_1\drt_2
\Eeq

The sum of all irreducible \dg\ gives the nontrivial part of the \DEQ\ -- the kernel ${\cal Q}$ of the integral operator in \Eq{DEQ}, known as mass operator. As it is seen by the definition of \DEQ, the mass operator ${\cal Q}$ renormalizes the free \pr\ $G_o$.

Let us proceed now with calculation of the scattered wave \I. Discrete \RM\ have a very good property: to each perturbation order corresponds a given number of \sc s. To describe the wave propagation through the disordered medium we need to calculate the {\it 4-points \CF\ (or \CF\ of two \pr s)}\ defined by:
\def\Gam#1#2#3{\Gamma_{#1}(\rt_{#2},\rt^*_{#2};\rt_{#3},\rt^*_{#3})}
\beq
\Gam{}{}0=\av{\Gr{}0\, G^*(\rt^*,\rt_0^*)}
\eeq
Let two point-like wave sources are given in a medium. We are interested in the wave amplitudes for two arbitrary fixed points, so we intend to calculate the 4-points \coh\ \fu. Each wave, passing through a sequence of \sc s generates partial waves from various intermediate states, and they may interfere.
Besides, during the \sc\ in the intermediate states some amplitudes, coming from one source may add to these descending to another, and these intermediate interferences are not included into the \AGF. The energy transfer is accounted in the \CF\ of \pr s and 4-points \dg\ (Fig.~\ref{BSeq}). It is represented by dashed lines connecting upper and lower solid lines. Because of that the 4-point correlator is not a product of two \AGF s, but is the averaged product of two \RGF s. The \eq\ between the 4-points \fu, the \AGF, and the \CF\ of the scatterers (the {\it \IC}) is called \BSEQ:
\Wbeq[BSEQ]
\Gam{}{}0=\Gam0{}0+\!\int\!\Gam0{}1\,{\cal K}(\rt_1,\rt^*_1;\rt_2,\rt^*_2)\,
\Gam{}20 \drt_1\drt^*_1\drt_2\drt^*_2\:,
\Weeq
where $\Gamma_0\!:=\!\av G \av{G^*}$ and the kernel ${\cal K}$ of the integral operator of the \BSEQ\ is the sum of all irreducible \dg.
\def\Fpoints{
\multiput(0,0)(0,20)2{\multiput(0,0)(15,0)2{\PTL}}
\put(0,-4){\mbrt{^*_1}}\put(15,-4){\mbrt{^*_2}}
\put(0,24){\mbrt{_1}}\put(15,24){\mbrt{_2}}
}
\def\KER{\put(0,0){\framebox(15,20){{\Huge ${\cal K}$}}} \Fpoints }
\unitlength=0,8mm \lth{0.4pt}
\bfigW
\begin{picture}(218,70)(-50,-6)
\put(-6,10){\mb{$b)$}}
\put(0,0){\KER}
\put(22,10){\mb{{\large $:=$}}}
\multiput(29,0)(43,0)2{\Vdash\PTL}
\multiput(29,-4)(43,0)2{\mb{$\rt^*_1\!\!\equiv\!\rt^*_2$}}
\multiput(29,24)(19,0)2{\mb{$\rt_1\!\!\equiv\!\rt_2$}}
\multiput(36.0,10.0)(24,0)4\PLUS
\put(48,0){\ArC}
\put(48,0){\Vdash\PTL}
\multiput(40,0)(49.2,0)2{\vector(1,0){15.2}}
\put(40,-4){\mbrt{_1^*}}
\put(56,-4){\mbrt{_2^*}}
\put(48,-4){\mbrt{'^*}}
\put(72,20){\ArC}
\multiput(80,20)(25.8,0)2{\vector(-1,0){15.2}}
\put(64,16){\mbrt{_1}}
\put(80,16){\mbrt{_2}}
\put(72,22){\mbrt{'}}
\put(90,0){\Fpoints}
\multiput(90.3,0.4)(0.9,1.2){16}{\PT}
\multiput(104.7,0.4)(-0.9,1.2){16}{\PT}
\put(115,8){\Huge\dots}
\put(0,38){
\put(-6,10){\mb{$a)$}}
\multiput(0.5,0)(0,20)2\BOX
\put(0,-4){\mbrt{^*}}
\put(0,24){\mbrt{}}
\multiput(0,0)(99,0)2{
\put(5,0){\framebox(15,20){{\Huge $\Gamma$}}}
\put(1,0.1){\vector(1,0){23.5}}
\put(24.5,20.1){\vector(-1,0){23.5}}
\multiput(25,0)(0,20)2\BOX
\put(25,-4){\mbrt{_0^*}}
\put(25,24){\mbrt{_0}}
}
\put(55,-4){\mbrt{_0^*}}
\put(55,24){\mbrt{_0}}
\put(32,10){\EQUAL}
\multiput(40,0)(30,0)2{
\multiput(0,0)(0,20)2\BOX
\multiput(-0.8,0.1)(0,20)2{{\lth{2pt} \line(1,0){13}}}
\put(6.4,9.6){\mb{\Huge $\Gamma$\Large $\!_0$}}
\put(0,-4){\mbrt{^*}}
\put(0,24){\mbrt{}}
\put(7,0.2){\mb{$\blacktriangleright$}}
\put(7,20.2){\mb{$\blacktriangleleft$}}
}
\multiput(54,0)(0,20)2\BOX
\put(62,10)\PLUS
\put(84.3,0){\KER}
}
\end{picture}
\caption{\mbox{a)\,\BSEQ\,;\ \ b)\, Kernel ${\cal K}$ as a sum of all irreducible \dg.}}
\label{BSeq}
\efigW

If we know the distribution of the wave sources, we may proceed to the final quantity of interest -- the {\it \CF\ of the scattered field}\ :
\Wbeq
\Gamma(\rt,\rt^*)=\av{E(\rt)\,E^*(\rt^*)}=
\int\Gam{}{}0\,j(\rt_0)j^*(\rt_0^*)\,\drt_0\drt^*_0\:.
\Weeq

Both \eq s \Eq{DEQ} and \Eq{BSEQ} in the theory of wave propagation in disordered media are exact \eq s and their analytical solution in the general case is not known. For this reason, one natural question arises: how to do proper \app s in the mass operators ${\cal Q}$ and ${\cal K}$\ ? Usually, the common adopted \app s do not restrict to given order of the perturbation, but to the type of irreducible \dg\ taken into account.

\subsection{The characteristic scales.}
Let us briefly discuss the characteristic space-time scales, used in this study. The first scale is set by the \TMFP\ and the mean time between two \sc s. The variables, which vary in intervals of the order of this scale will be denoted by capital letters. The \IC\ decreases rapidly with the distance $\M{\rs_1\-\rs_2}$, so it determines another characteristic scale, in which particle radius, time for passing the particle size and life-time of the excitations are included. The variables, which change inside that scale, will be denoted by small letters. When we calculate quantities of interest, we face with integration over distances from $-\infty$ to $+\infty$. In this case some additional \app s may be done under the integral. Nevertheless, we should treat the sign $\infty$ not as an absolute symbol, but as a quantity, attached to one of the above mentioned scales. Such a case is realized when we assume the \FrA\ to be valid under the integral.

Note that the large scale imposes a restriction to the medium size. For example, a \sc\ slab must not be smaller than the \TMFP\ \cite{McKSJ}. Moreover, the incident wave packet must not have a \coh\ length, smaller than the total mean free path.

In what follows, we introduce large scale coordinates
$$\Rv:=\frac{\rs\+\rs^{\,*}}2\;\;\;\;T:=\frac{\ttt\+\ttt^*}2$$
and relative coordinates
$$\rv:=\rs\-\rs^{\,*}\;\;\;\;t:=\ttt\-\ttt^*\;.$$

Finally, we would like to draw attention to another important question, related to the two characteristic space-time scales. In general, the \IC\ depends dominantly on the spatial separation $|\rs_1\-\rs_2|$. When this separation is comparable to the characteristic length of the large scale, the \IC\ tends to zero. In contrast, in the case of small separations $|\rs_1-\rs_2|$, close to the characteristic length of scale of microstructures, the correlators $\psi_{\ep}$ give a considerable contribution to $\Gamma$ in different perturbation orders. However, only \ldg\ expansion of $\Gamma$ (Fig.~\ref{LCdg}(a)) contains the \IC s with such small magnitudes of $|\rs_1\-\rs_2|$. That's why, the principal contribution to the 4-points \CF\ is given by the \ldg.

\def\PO#1{\multiput(0,8)(4,0){#1}{\circle*{0.8}} }
\unitlength=0.7mm \lth{0.3pt}
\bfigW
\begin{picture}(160,60)\put(10,35){
\put(-7,8){\large a)}
\BX L\PTL\PTL
\put(23,10)\EQUAL
\multiput(30,0)(27,0)2{\Vdash\PTL}
\multiput(37,10)(27,0)2\PLUS
\put(44,0){\QQ\PTL}
\multiput(71,0)(13,0)2{\QQ\PTL}
\put(97,0){\Vdash\PTL}
\multiput(104,10)(22,0)2\PLUS
\multiput(111,0)(77,0)2{\PO3}
\multiput(133,0)(28,0)2{\QQ\PTL}
\multiput(146,0)(28,0)2{\Vdash\PTL}
\multiput(149,-8)(0,20)2{\PO3}
\put(181,10)\PLUS
}
\put(10,5){
\put(-7,8){\large b)}
\BX C\PTL\PTL
\put(23,10)\EQUAL
\put(33,0){\Qq \multiput(0,0)(0,20)2\PTL \multiput(12.8,0)(0,20)2\PTL
\multiput(0.3,0.4)(0.78,\step)\points\PT \multiput(0.3,19.6)(0.78,-\step)\points\PT}
\multiput(52,10)(122,0)2\PLUS
\put(63,0){%
\multiput(0,0)(12.2,0)2\Qq
\put(12.7,0){\Vdash\PTL}
\multiput(0,0)(0,20)2\PTL \multiput(25.5,0)(0,20)2\PTL
\multiput(0.5,0.4)(0.94,0.74){27}\PT \multiput(0.5,19.6)(0.94,-0.74){27}\PT
}%
\multiput(98,10)(22,0)2\PLUS
\multiput(105,0)(77,0)2{\PO3}
\put(127,0){%
\multiput(0,0)(25.6,0)2\Qq
\multiput(13.1,0.4)(0.78,\step)\points\PT \multiput(13.1,19.6)(0.78,-\step)\points\PT
\multiput(0.5,0.4)(1.1,0.56){34}\PT \multiput(0.5,19.6)(1.1,-0.56){34}\PT
\multiput(0,0)(12.8,0)4{\multiput(0,0)(0,20)2\PTL}
\multiput(15,-8)(0,20)2{\PO3}
}%
}
\end{picture}
\caption{(a) Ladder \dg. (b) Maximally crossed \dg.}
\label{LCdg}
\efigW

On the other hand, if the disordered medium has a {\it\TRS\ }, the \cdg\ (Fig.~\ref{LCdg}(b)) may be converted to ladder type \dg\ by a time-reversal operation \cite{LN,Gb,McKSJ}. Therefore, for a time-reversal invariant medium, the \cdg\ give a significant correction (in order $1/k\el$) to $\Gamma$. These \dg\ describe the experimentally observed effect of \CBSC\ of light. It must be pointed out here, that this effect of \I\ amplification at backward direction is due only to the \TRS\ of the \sc\ medium, but not to the statistics of the impurities \cite{Gb}. The effect of \CBSC\ may be described physically by the constructive interference of a given \MSC\ path with itself but time-reversed. If the medium does not have a \TRS, the \cdg\ can not be represented as time-reversed \ldg. That's why, in the case of broken \TRS, their contribution to $\Gamma$ becomes negligible.

\subsection{Coherent potential \app.}
Let us now determine the \AGF. This requires to fix the \app\ for the integral operator ${\cal Q}$ in the \DEQ. In addition we suggest that the average over the disorder $\tilde\ep$ restores translational symmetry, so our quantities of interest will depend on the spatial separation $|\rs\-\rs'|$. Commonly accepted  approach is the simplest one, called {\it \coh\ potential \app\ }\ \cite{McKSJ}:
\beq\begin{array}{rl}
\DS12:=&\Go12\Fi{\rs_1\-\rs_2;\ttt_1\-\ttt_2}\;\\
\Fi{\rs_1\-\rs_2;\ttt_1\-\ttt_2}:=&k_0^4\PE12\:,
\end{array}\eeq
where $\Phi$ is the \IC\, multiplied by $k_0^4$. This first order \app\ for ${\cal Q}$ has the following diagrammatic \rep :\\

\unitlength=1.0mm \lth{0.4pt}
\begin{picture}(80,10)(-25,-5)
\put(0,0){\makebox(0,0){{\large ${\cal Q}$}\ {\large $\;\;\approx$} } }
\put(10,0){\COR012 \line(1,0){16}}
\end{picture}

Such assumption for ${\cal Q}$ leads to the following \FI\ of \DEQ\ :
\beq[DEQ1]
\GmK{}^{-1}=\GoK{}^{-1}\!\!-\DSK{}\;\;,
\eeq
where
\beq[Q]
\DSK{}=(2\pi)^{-4}\!\!\!\int\!\GoK'\Fi{\kv\-\kv',k_0\-k_0'}\D{k}'\d k_0'\;\;.
\eeq

\subsection{White-noise \app.}

The \IC\ is determined by the size and motion of the scatterers. Furthermore, it may depend on the interactions between the \sc\ particles. The simplest assumption for the \IC\ is a the {\it white-noise \app}. It means:

1/ low density of the impurities;

2/ uncorrelated disorder -- no interactions between the \sc\ particles;

3/ point-like scatterers;

4/ nonrelativistic motion of the impurities;

5/ structureless \sc\ particles -- not any internal excitations;

6/ absence of any \ic\ quantum effects, like Compton effect, etc.

For low densities and for noninteracting particles the \CF\ of scatterers is proportional to the density of \sc\ particles and has width, approximately equal to the particle radius \cite{McKSJ2}. The essence of the white-noise model consists in the following: for \sc\ particles of size, $a$,much smaller than the wavelength of light ($a\!\ll\!\lambda$), the Fourier-transformed correlator $\Phi(\vec q,t)$ will be almost independent of the \sc\ transfer vector $\vec q$, and hence the \sc\ will be isotropic. Besides, during the field transport from one point to another within the medium, all transfers of the intermediate wave vectors conserve their magnitudes $\M{\!\kv_j\!}=\!k_0$.

To proceed further, we first must define the \TMFP\ $\el$, which plays an important role as a basic scale unit. It is given by:
\beq[el]
\frac1\el:=\frac1{\el_i}+\frac1{\el_e}\;\;,\;\;\;\;
\el_i:=\frac1{k_0\ep''}\;\;,\;\;\;\;\el_e(\o{}):=
\frac{4\pi c^4}{n\ep_*^2\o{}^4}\;,
\eeq
where $\el_i$ is the \IMFP, due to the absorption processes, $\el_e$ is the \EMFP, $n$ is density of impurities and $\ep_*$ is polarizability of scatterers.

For \st\ white noise model the correlator $\psi_{\ep}$ is a delta \fu\ in spatial variables, normalized to $4\pi\!/\el_e k_0^4$ \cite{McKSJ}. Therefore the \FI\ of $\Phi$ becomes $4\pi\!/\el_e\,\DE(k_0\-k_0')$ and one can obtain for \Eq{Q} the result $ik_0/\el_e$. Replacing it in \Eq{DEQ1} and using the \FI\ of unperturbed \GF :
\def\kprop#1{\frac1{k^2\-k_0^2\!\mp\!ik_0/\el_{#1}}}
$$G_0^{+/-}\kko{}=\kprop{i}\;,$$
we obtain the \AGF:
\beq[Gk]
\av{G^{+/-}\kko{}}= \kprop{} \;.
\eeq
Here the subscripts $"+"$ and $"-"$ denote advanced and retarded \pr s.

Let us now consider the time-dependent white-noise model. In this case the scattered \I\ is related to the \FI\ of the \CF\ \cite{St}:
\beq
\Phi(\qv,t)=\frac{4\pi}{\el_e}\exp\br{-a(t)\,\qv^2\,}
\eeq
where $\qv$ is the transfer wave-vector and $a(t)$ describes the character of the impurities motions. The effect of particle dynamics on the \mul\ \sc\ of light has been discussed in \cite{Gb} and \cite{St} for two special cases:

(1) the scatterers have a Maxwell-Boltzmann velocity distribution; and

(2) the scatterers are themselves diffusing in the medium with a given
diffusion constant.

In our further investigations we shall not specify the form of the correlator $\Phi(\qv,t)$. In general, we assume nonrelativistic motions of scatterers $\sqrt{<\!\!v^2\!\!>}/c\ll 1$. Then the frequency change during \sc\ is negligible $\Delta\o{}/\o0\!\ll\!1$ and we can consider the propagation of an almost-monochromatic wave with frequency close to the frequency of the incident light $\o0$. According to this assumption $\Phi$ will not depend in an important way on $a(t)$. Then the \FI\ of ${\cal Q}$, \ex \Eq{Q}, coincides with the \st\ one, after the formal replacement $k_0\rightarrow\o{}/c$. Therefore the \AGF\ in $(k,\o{})$-\rep\ gets the same form as \Eq{Gk}. The propagation of a scalar wave in an infinite medium is described by the retarded \GF:
\beq
D(\rs\-\rs',\o{})=\frac{\exp\Br{i\frac{\o{}}{c}-
\frac1{2\el(\o{})}}\!|\rs\-\rs'|}{4\pi|\rs\-\rs'|}
\eeq
\vspace{1em}
where $D\!:=\!\av{G^-}$.

\section{The sum of ladder diagrams.}
In the \WS\ limit $\lambda\!\ll\!\el$ the leading \app\ to the 4-points \CF\ $\Gamma$ is given by the sum of \ldg, represented in (Fig.~\ref{LCdg}(a)).

The \lpr\ describes the \ICMSC\ of light through the medium, i.e. the ordinary diffusion in the medium. To obtain the sum of \ldg, we must include only the ladder term in the expansion of ${\cal K}$. The \BSEQ\ then reduces to an \ieq\ for the \lpr. Further we will use large scale and relative coordinates, introduced in section II\,B . The \pr\ $L$ is most easily calculated in $(\kv,\o{})$-space. The \FI\ of the sum of \ldg\ then satisfies the \ieq:
\WBeq[31]
\lefteqn{ L (\KvO \,; k_0\Sv_1,\o1\,; k_0\Sv_2,\o2)=
\Fi{k_0\SS12 ,\,\oab}+ } \NN
& &+(2\pi )^{-4} \int\!\Fi{k_0\Sv_1\-\kv ,\,\o1\-\o{}} \DKO
L (\KvO \,; k_0\Sv_1,\o{} \,; k_0\Sv_2,\o2)\d\kv \d{\o{}}\;,
\WEeq
where $\O{}$ is the frequency change during \sc, and  $\Sv_1$, $\Sv_2$ are unit vectors.

In the \WS\ regime the \FrA\ is allowed:
\Wbeq
D(\Rr+{},\Oo+)\approx\, \exp\br{ik_0\Sv_{R}\.\frac{\rv }{2}}\, D(\Rv,\Oo+)
\;\;;\;\;\Sv_R := \U{\Rv}
\eeq
According to this \app\ the product of retarded and advanced \GF s becomes:
\beq[33]
\DKO = (2\pi)^3 \int\!\DE(\kv -k_0\Sv_{R})\e^{-i\Kv\.\Rv}\dRO\dR
\eeq
Replacing \Eq{33} in the \ieq\ \Eq{31}, we get:
\Beq[Lko]
\lefteqn{ L (\KvO;\, k_0\Sv_1,\o1;\, k_0\Sv_2,\o2)=
\Fi{k_0\SS12 ,\,\oab}+ } \NN
& &+(2\pi )^{-1} \int\!\Fi{k_0\SS1R ,\,\o1\! -\!\o{}}\,
\e^{-i\Kv\.\Rv}\,Q(\Rv,\O{},\o{})\,
L (\KvO;\, k_0\Sv_1,\o{}; \, k_0\Sv_2,\o2) \dR \d{\o{}}\;,
\Eeq
where
\beq
\nonumber
Q(\Rv ,\O{},\o{}):=\dRO =\frac{\exp\Br{i\frac{\O{}}{c}-
\frac1{2\el(\Oo+)}-\frac{1}{2\el(\Oo-)}}R}{(4\pi R)^2}
\Weeq
For small frequency change on \sc\ $\O{}\!\ll\!\o{}$ \ex\
\Eq{el} for $\el$ becomes:
\beq
\el^{-1}\br{\Oo\pm} \approx \el_i^{-1}\!+\el_e^{-1}(\o{})\br{1\!\pm\!2\frac{\O{}}{\o{}}}
\eeq
and
\beq
Q(R,\O{},\o{})\approx (4\pi R)^{-2}
\exp\Br{\br{ i\frac{\O{}}{c}-\frac{1}{\el (\o{})}} R}
\eeq

In the \st\ case the correlator $\Phi$ turns out to be a $\delta$-\fu\ with respect to $\o{}$, and in the non\st\ case the last one is almost $\delta$-\fu. This circumstance afford us to take {\bf $Q(\Rv,\O{},\o{})\!=\!Q(\Rv,\O{})$} in the integrand of \Eq{Lko}. So without any loss of generality we get $L(\KvO;\,k_0\Sv_1,\o1;\,k_0\Sv_2,\o2)\!=\!L(\KvO;\,k_0\Sv_1,k_0\Sv_2;\oab)$ and perform the \Ftr\ with respect to $\o{}$ ($\o{}\!\rightarrow\!t$) in \Eq{Lko}. Taking into account that $L$ satisfies \eq\ \Eq{Lko}, we derive the relation between the $(n\+1)^{\text{th}}$ and $n^{\text{th}}$ generic terms of the sum $L$:
\def\LKOt#1{L_{#1}\br{\KvO,t}}
\def\LKOkt#1#2#3{L_{#1}\br{\KvO;\,k_0\Sv_{#2},k_0\Sv_{#3};\,t}}
\def\Fikt#1#2{\Fi{k_0\SS{#1}{#2},t}}
\Wbeq
\LKOkt{n+1}12=\int\!\Fikt1R\,\e^{-i\Kv\.\Rv}\,Q(\Rv,\O{})\,\LKOkt{n}R2\dR
\Weeq


Let us assume that {\it s-wave \app}\ is valid, i.e. the {\it bare diffusion \pr} describes propagation of a spherical wave. It means that $\LKOkt{n}12$ is independent on the directions:
\beq[Sw]
\LKOkt{n}12 \approx \LKOt{n}\;,
\eeq
where
\beq
\LKOt{n}\!:=(4\pi)^{-2}\!\!\!\int\!\!\LKOkt{n}12 \dS_1\dS_2\:.
\eeq

Note that {\bf for the \st\ case in white-noise model eq.\Eq{Sw} becomes an exact \eq}\ as far as the \FrA\ has already been assumed. For the non\st\ case {\it the s-wave \app}\ follows from the \FrA\ and from {\bf the quite weak angular dependence of the \IC}. In some papers \cite{St,Gb} it was related to the \DA, but in our opinion it follows from {\bf the \WS\ \app}, which permits doing the \FrA. As we will see bellow {\bf the \WS\ does not require the \DA}.

Let us define the angular averaged correlator \cite{Gb}:
\beq
\gamma f(t):=\int\!\Fikt12\frac{\dS_2}{4\pi}\;\;,
\;\;\;\;\;\gamma :=\frac{4\pi}{\el_e}\;.
\eeq
Then each term fulfills the relation:
\beq
\LKOt{n+1}\!=\gamma f(t)Q(\KvO)\LKOt{n}\,,\;\;n=1,2,\dots
\eeq

From this point on the single-\sc\ term will be explicitly separated:
$$L = L_1+\LL \;\;,\;\;\;\;\LKOt1:=\gamma f(t)$$

\beq
\Lt{\KvO}=\gamma^2 f^2(t) Q(\KvO)+\gamma f(t)Q(\KvO)\Lt{\KvO}
\eeq
where the \fu\ $Q(\KO)$ has the following form:
\beq
\begin{array}{l}
Q(\KO)= \frac{1}{8\pi K}[\atgKl{+} +\atgKl{-}]+\\
\\
+\frac{i}{16\pi K} \{ \lnKl{+} -\lnKl{-} \}
\end{array}
\eeq

Furthermore, the sum of \ldg\ may be written as a geometric series in the variable $\gamma f(t)Q(\KO)$ :
\Beq[GS]
\begin{array}{rl}
\Lt{\KO}=&\sum_{n=1}^{\infty} \Ln{\KO}\\
& \\
\Ln{\KO}:=&\gamma f(t)(\gamma f(t)Q(\KO))^n\;.
\end{array}
\Eeq
For nonrelativistic motions of scatterers, the frequency change on \sc\ is negligible, i.e. $\O{}/\o0\ll 1$, so we can replace $\O{}=0$:
\def\Atg#1{\frac{\arctg({#1})}{{#1}}}
\def\AKl{\Atg{K\el}}
\beq
\gamma Q(K):= \gamma Q(K,0)= \frac{\el}{\el_e}\,\AKl
\eeq
and the $(n\+1)^{\text{th}}$ ladder term gets the form:
\beq[Kn]
\begin{array}{rl}
\Ln{K}=&\frac{4\pi}{\el}\gt\br{\gt\frac{\arctg(K\el)}{K\el}}^{\!\!n} \,,\\
\gt:=&\frac{\el}{\el_e}f(t)=\frac{f(t)}{1+\mu}\leq 1\;.\\
\end{array}
\eeq

On taking the \Ftr\ with respect to $\Kv$, we obtain:
\def\iLn{\InT\e^{i\fRl p}\br{\frac{\arctg p}p}^{\!n}p\d{p}}
\def\PIterm#1{\INT\frac{\;#1}{(2u)^{n-1}}
\br{\ln^{\!2}\!\!\UPM\+\pi^2}^{\!\!\frac{n}2}
\sin\br{n\,\arctg\frac{\pi}{\LP}}\du }
\def\Pterm#1#2{ \frac{g^{n+1}(t)}{\pi{#2}} \PIterm{#1} }
\beq[LnI]
\Ln{R}=\frac{g^{\!n+1}(t)}{i\pi\Rl}\iLn
\eeq
Evaluation of the integral \Eq{LnI} in the complex plane is described in Appendix \Eq{A1}. The result is :
\Wbeq[LnR]
\Ln{R}=\Pterm{\;\E{R}}{\Rl}
\Weeq

This is the final analytical form for $\Ln{R}$, we can reach in the case of $s$-wave \app. The \ex s for $L_2(R,t)$,\dots, $L_7(R,t)$ are given in Appendix \Eq{A2}).

The geometric series \Eq{GS} are convergent for all $K$, since $0\!\leq\!\gt\AKl\!<\!1$. Then the sum of \ldg\ (without single-\sc\ term) in wave-vector space is:
\beq[LK]
\Lt{K}=\frac{4\pi}\el\frac{g(t)\AKl}{\GT-\AKl}\;,
\eeq
and the \lpr\ in coordinate space is given by:
\beq[LRv]
\Lt{\Rv}=\frac{g(t)}{2\pi^2\el}\int \frac{\e^{i\Kv\.\Rv}\;\AKl}{\GT-\AKl}\d^3\Kv\;.
\eeq

Let us now define \EDL\ by the transcendent \eq\ \cite{Dev}:
\beq[EDL]
\CPM =\GT\ \;\;\;\;\;\gt\leq 1
\eeq

The \EDL\ (in units of total mean free path $\el$) is defined as the reciprocal value of the root of \eq\ \Eq{EDL} -- $\chi^{-1}$. The quantity $\chi^{-1}\+1$ represents the {\it effective order of \sc}. We have to point out here that the \EDL\ does not impose any rigid restriction to the long diffusion paths, i.e. there exist diffusion paths greater than $\chi^{-1}$ even in the case of very large effective diffusion lengths. The magnitude of $\chi^{-1}$ influences only the speed of convergence of the infinite sum $\LL$.

In fact, the root of transcendent \eq\ \Eq{EDL} gives the two poles at $\pm i\chi$ in the integrand of \Eq{LRv}. Evaluation of integral \Eq{LRv} is carried out in the complex plane(see Appendix). The final result for the sum of \ldg\ (without single-\sc\ term) is:
\Wbeq[LRt]
\Lt{\Rv}=
\frac1\Rl\!\Br{ A(\xt{})\,\exp\br{-\frac R\el \xt{}}+\Iterm\GT{\E R} }
\;,\;\;\;\;\;A(\xt{}):=\frac{2\Xt{}}{(1\-\Xt{})^{-1}\-\GT}
\eeq

It is convenient to define an integral operator $\Fg$ by:
\beq[defFU]
\FU{\gt}{H(u,\dots)}:=A(\xt{})\,H(\xt{} ,\dots)\,+\,\Iterm{\GT}{H(u,\dots)}
\Weeq
where $H(u,\dots)$ is an uniformly continuous \fu. Then:
\beq[LF]
\Lt{\M\RaRb}=\FU{\gt}{\ERR\RaRb}
\eeq
The first term in \Eq{LRt} corresponds to the familiar diffusion \pr\ \cite{Dev} and {\bf the second term becomes significant when the contributions of higher \sc\ orders (or equivalently long diffusion paths) to $\LL$ are negligible}.

The $N^{\text{th}}$-order expansion of \Eq{EDL} is:
\Beq[sum1]
\CPM&\approx&\sum_{n=0}^{N}\frac{\chi^{2n}}{2n+1}\;,\NN
\frac1{2\chi^2}\br{\frac1{1\-\chi^2}-\CPM}&\approx&\sum_{n=1}^{N}\frac{n\chi^{2n-2}}{2n+1}
\Eeq

In the limits of large effective diffusion lengths $\chi\ll 1$, the first order \app\ (N=1) is acceptable:
\Beq[o1]
\GT\approx 1+\frac{1}{3}\chi^2\;&\Rightarrow&\;\;
\chi^2\approx\chi_1^2:=3\GT-3\:;\NN
&\Rightarrow&\;\;\frac1{2\chi_1^2}\br{\frac1{1-\chi_1^2}-\GT}\approx\frac13
\Eeq

For $N=1$ in the \st\ case ($f(t)=1$), without absorption ($\gt =f(t)$),
the exact result for the sum of \ldg\ follows:
\Wbeq[ldp]
\LL(R)=\FU{1}{\frac{\E{R}}{\Rl}}=\frac{1}{\Rl}\Br{\ 3+\Iterm{1}{\E{R}}}
\Weeq

\section{Boundary conditions for a semi-infinite medium.}
Up to now, we have obtained the sum of \ldg\ \Eq{LRt} in the case of an infinite \sc\ medium. However, our aim is to study the light reflected by a semi-infinite \sc\ medium occupying the half-space $Z\!>\!0$. In order to solve this task we must impose proper boundary conditions on the \pr\ $\LL$. It satisfies an \ieq, which has the following diagrammatic form:

\unitlength=0.7mm\lth{0.4pt}
\begin{picture}(92,24)(0,-1)
\BX\LL\BOX\BOX
\put(23,10)\EQUAL
\multiput(30,0)(30,0)2{\QQ\BOX}
\put(37,9){\mb{{\huge $L$}{\Large $_2$}}}
\put(44,0){\Vdash\BOX}
\put(74,0){\BX\LL\PTL\BOX}
\put(52,10)\PLUS
\end{picture}

The usual \app\ made is that the term $L_2$ is too small in the infinite sum $\LL$. By this way, the upper \ieq\ may be reduced to the Milne's \ieq:

\unitlength=0.7mm\lth{0.4pt}
\begin{picture}(85,30)(-20,-5)
\BX\LL\BOX\BOX
\put(23,10)\EQUAL
\put(30,0){\QQ\BOX}
\put(43,0){\BX\LL\PTL\BOX}
\end{picture}

When the \EDL\ $\chi^{-1}$ approaches infinity, the sum $\LL(K)$ is a geometric series with $Q(K)\!\lesssim\!1$, so that the contribution of $L_2(K)$ to the \sc\ \pr\ is truly small. This situation corresponds to the case of \st\ macro scatterer, without absorption losses in it. However, when the \EDL\ decreases significantly and becomes close to 1 (\EDL $\:\approx\,$\EMFP), then in accordance with the strong convergence of $\LL$, the contributions of lower \sc\ orders (in particular $L_2$) become important. That's why in this case, reducing the \ieq\ for $\LL$ to the Milne's one brings a small error.

The integral Milne's \eq\ has the advantage to be solved exactly for point-like scatterers, by mean of the Wiener-Hopf method \cite{Dev,PS}. It can be shown that very far from the surface of the medium ($Z\!\!\gg\!\el$\,) $\LL$ obeys a diffusion \eq\ \cite{PS,Dev,McKSJ} with the boundary condition $\LL(-Z_{as})\!=\!0$\ . It means that the diffusion \pr\ (the asymptotic form of $\LL$ when $R\!\rightarrow\!\infty$) vanishes on a trapping plane, located at an extrapolation distance $-Z_{as}$. The computation of $Z_{as}$ was carried out with high precision \cite{PS}:
$$\Mas:=\frac{Z_{as}}\el=0.710\,446\;.$$

It should be noted, that the asymptotic boundary condition proposed above gives a correct solution of the Milne's \eq\ only in the \DA. However, when we are concerned with the \coh\ \I\ for large \sc\ angles, we should correct the extrapolation distance $Z_{as}$ \cite{AWM}. The exact solution of Milne's \ieq, beyond the \DA, has been obtained by Mark \cite{Mark}.Imposing the boundary condition for a semi-infinite medium on it, we obtain the correct location of the trapping plane as a root of an algebraic \eq\ (see appendix \ref{}). In our further calculations of scattered \I\ we shall adopt the exact Milne's number:
$$M:=\frac{Z_0}\el=0.689\,710\;,$$
which gives the effective boundary for the case of infinite \EDL.

As the explicit form of the \lpr\ for an infinite \hom\ medium has already been found, we could obtain $\LL$ for a half-space, using the exact boundary condition and the well known method of images:
\Wbeq[bc1]
\Lt\Rab:=\int\!\Lt\Kv\Br{\e^{i\Kv\.(\Rv_1\!-\Rv_2)}-\e^{i\Kv\.(\Rv_1\!-\Rv_2^*)}}
\frac{d^3\Kv}{(2\pi)^3}=\Lt\RaRb-\Lt{\RaRb^*}\;,
\Weeq
so that
$$\Lt\Rab=0\;\;\;\text{as}\;\;\;\;\Rv_2=\Rv_2^*\;,$$
where the images are located at
$$\Rv_2:=(X_2,Y_2,Z_2)\,,\;\;
\Rv_2^*:=(X_2,Y_2,Z_2^*)\,,\;\; Z_2^*:=-Z_2-2Z_0\;.$$
In conclusion it may be stressed that the main differences between the \DA\ and our approach consist in:
\ \\

{\bf
\ 1/The way, in which the \pr\ $\Lt\RaRb$ is defined eq.\Eq{LRt}.

\ 2/The value of the parameter $M$.} \\

\section{Incoherent multiple scattering intensity.}
We have  obtained the sum of \ldg\ \Eq{LRt}, which characterizes the \sc\ medium. The distribution of the wave sources is known. Then we may proceed to the calculation of the part of the time-dependent \CF, describing the \mul\ \ICSC. It is  defined by:
\WBeq[e54]
\lefteqn{\GL (\Rv,\rv\,;\:T,t):=\int\!\! D\br{\RR_1 +\rr1 ,\,\TT1 +\tt1}
D^*\br{\RR_1 -\rr1 ,\,\TT1 -\tt1 }\times } \NN
& & \times\,\LL \br{\Rab ,\rab\,;\:T_1,T_2,t_1,t_2}\,
E\br{\Rr+2,\Tt+2} E^*\br{\Rr-2,\Tt-2}\dR_1\dR_2\dr_1\dr_2\,\dT_1\dT_2\dt_1\dt_2
\Eeq
Since the medium is invariant under translations of \ $T$ \ we have:
\beq
\LL(\Rab,\rab\,;\:T_1,T_2,t_1,t_2)=\LL (\Rab ,\rab\,;\:T_1\-T_2,t_1,t_2)
\eeq
If we assume that the fields are monochromatic waves $E(\Rv ,t)\!:=\!E(\Rv )\,\eot$, after \Ftr\ of the \fu s under the integral \Eq{e54} with respect to the time variables, we obtain:
\Beq
\lefteqn{\GL(\Rv,\rv\,;\:T,t)=\GL\Rrt =\eot\int\!\e^{i(\o{}+\o0\!)t}\,
D\br{\RR_1+\rr1 ,\o{}} D^*\br{\RR_1-\rr1 ,\,\o{}}\times } \NN
& & \times\,\LL (\Rab ,\rab ;\,\o{},-\o0)\,E\br{\Rr+2}\, E^*\br{\Rr-2}\,
\doo\dR_1\dR_2\dr_1\dr_2
\Eeq
The integration over $\dr_1 ,\,\dr_2$ and $\doo$ gives the \FI\
$\Lt{\Rab ,k_0\Sv_{RR_1},k_0\Sv_{R_2}}$, so the field \CF\ becomes:
\beq
\GL\Rrt = \frac{\eot }{16\pi^2}\int\!\exp(ik_0\Sv_{RR_1}.\!\rv )\,\DDR1\,
\Lt{\Rab ,k_0\Sv_{RR_1},k_0\Sv_{R_2}}\,\M{E(\Rv_2)}^2\dR_1\dR_2
\eeq
Taking into account that in $s$-wave \app\
$\Lt{\Rab ,k_0\Sv_{RR_1},k_0\Sv_{R_2}}\!=\!\Lt{\Rab}$, after \Ftr\ we get:
\beq[G1]
\GL\Rkt:=\int\!\e^{-i\kv\.\rv}\,\GL\Rrt\dr
= \frac{\pi\eot }{2}\int\!\DE(\kv -k_0\Sv_{RR_1})\,\DDR1\,
\Lt{\Rab}\M{E(\Rv_2)}^2\dR_1\dR_2
\Weeq

The \CF\ of the wave field $\GL\Rkt$ describes the \I\ of the \ICMSC\ scalar wave, which has a wave vector $\kv$ in the point $\Rv$. The \I\ in this point, for a given direction $\Sv$, can be obtained after integrating over the wave vector magnitude.
\beq[JJL]
J^{\LL}(\Rv,\Sv;t):=2\int_0^{\infty}\! k^2\,\GL\Rkt\,{\rm d}k
\eeq
In the case of semi-infinite medium we have a cylindrical symmetry. Z-axis is chosen to be normal to the surface and is orientated inside the medium. The other two axes can be orientated in such way, that:
\beq
\Sv:=\frac{\kv}{k}:=(\SIA,0,\COA)\;\;\;\;\;\;
\alpha:=\angle (\OZ,\Sv)
\eeq
According to the translation symmetry with respect to X and Y directions, without any limitation of generality we may choose $\Rv:=(0,0,Z)$. Then
\def\aRR{\sqrt{X_1^2+Y_1^2+(\ZZ_1)^2}}
\def\DEks#1{\DE(k\Sv - k_0\Sv_{RR_{#1}})}
\def\Dk{\DE(k\-k_0)}
\def\DELTA#1{
\DEks{#1}=\left\{
\begin{array}{ll}
\frac{(\ZZ_#1)^2}{k_0^2\!\MC^3}\,\Dk\;
\DE(X_#1\+\tga (\ZZ_#1))\,
\DE(Y_#1)\,
\TZ#1
&;\;\;\alpha\neq\pi/2 \\
\ \\
\;\;\;\;\;\;\;\frac{X_#1^2}{k_0^2}\;\Dk\,\DE(Y_#1)\,\DE(\ZZ_#1)\,\theta(-X_#1)
&;\;\;\alpha=\pi/2
\end{array}
\right.
}
\WBeq[d1]
\DELTA1
\Eeq
and
\beq
\JLZ\frac{\pi\eot}\MC \int\!\TZ1\EZ1\,|E(\rov,Z_2)|^2\,
\Lt{-\!\tga(\ZZ_1),0,Z_1;\,\rov,Z_2}\,\rov\dZ_2\dZ_1\;,\\
\eeq
where $\theta(X)$ is the Heaviside step \fu\ and $\Rv_2:=(\rov,Z_2)$. Using the cylindrical symmetry and the image boundary condition \Eq{bc1}, the following \ex\ is obtained:
\Beq[GL]
\lefteqn{\JLZ\frac{\pi\eot}\MC \Int_{\! Z_1=0}\!\!\TZ1\EZ1\Int_{\! Z_2=0}\!
\Int_{\!\rho =0}\!\int_{\!\phi =0}^{2\pi}\!\M{E(\rov,Z_2)}^2\!\times }\NN
& &\!\times\Br{\LL (\SQa\,,\: t)-\LL (\SQb\,,\: t)}\;
\rho\dro\dfi\dZ_2\dZ_1\;.
\Eeq

Referring to the result \Eq{LRt} and assuming, that the input wave is a plane wave and its $k$-vector is oriented to $\OZ,\;\;(\;\kv_0\!=\!(0,0,k_0)\;)$ i.e. $E(\Rv):=E_0\exp(i\kv_0\.\Rv -\frac{Z}{2\el})$ for $Z\geq 0$, it follows that:
\Beqn
\lefteqn{\JLZ\frac{2\NORM}\MC\,\Int_{Z_1=0}\TZ1\,\EZ1\,\times}\\
& &\times\Int_{Z_2=0}\Int_{\rho=0}\BR{
\FU{\gt}{ \frac{\E{\SQa}}{\SQa\:\el^3} } -
\FU{\gt}{ \frac{\E{\SQb}}{\SQb\:\el^3} } }
\,\rho\dro\dZ_2\dZ_1\;.
\WEeqn

Now let us introduce dimensionless variables:
\Beq
a:=\frac{Z_1}{\el}\;\;\;b:=\frac{Z_2}{\el}\;\;\;\;
h:=\frac{Z}{\el}\;.
\Eeq
\ \\
Since $\Fg$ is a linear integral operator, it may be written as
\beq[FL]
\JL 4\NORM\,\Fg [\HL{u}],
\eeq
where $H^{\LL}$ is defined by

\WBeq
\lefteqn{\HL{u}:=\frac{1}{2\MC}\,\Int_{a=0}\!\Theta\br{\frac{h-a}{\COA}}
\exp\br{\frac{a-h}{\COA}}\Int_{b=0}\e^{-b}\times } \NN
& &\times\Int_{\rho =0}\Br{\EXF\SQc -\EXF\SQd}\:\rho\dro\,\db\da
\Eeq
This integral can be calculated analytically and the final result is:
\beq[HL]
\HL{u}=\frac{\e^{-uh}\-\TECA }{(1\+u)(1\-u\COA )}\!
\br{\frac{\COA }{\COA\- 1}+\frac{1\-\e^{-2Mu}}{2u}}\+
\frac{\e^{-uh}\! -\!\e^{-h}}{1\-\COA }\,.
\eeq
Here $h$ is the distance from the surface to the point, at which the \I\ is measured, in units of \TMFP. Substituting \Eq{HL} in \Eq{FL} we get our final result for the \ICMSC\ \I:
\beq[JL]
\JL 4\NORM \Br{\Dterm{\GT}{\HL{\xt{}}}+\Iterm{\GT}{\HL{u}}}
\eeq
The connection between $\xt{}$ and $g(t)$ was defined by \Eq{EDL}.

For the stationary case ($f(t)\!=\!1$) without absorption ($g(t)\!=\!f(t)$) from \Eq{FL} and \Eq{JL} it follows: \beq[JLS]
\JL 4\NORM \FU1{\HL{u}}= 4\NORM \Br{3\,\HL{0} +\Iterm1{\HL{u}}}
\Weeq

It is easy to obtain the single \sc\ \I\ following the proposed analytical procedure -- one must simply exchange the \pr\ $\LL$ under the integral \Eq{G1} with the single \sc\ \pr\ (see Appendix 4.).

\section{Time reversal invariance and multiple coherent intensity.}
For a medium invariant under time-reversal all \cdg\ may be expressed by ladder type \dg\
with exception of the first one. The relation between the two types of \dg\ is:

\Wbeq
C\br{\Rab;\rab}=
\LL\br{\RbR1+2 +\rbr12,\RbR1+2 -\rbr12;\,\RaR1-2 +\rar12,\RaR2-1 +\rar12}
\Weeq

The \CSC\ \pr\ may be replaced by the \pr\ corresponding to the \ldg, but with variables as determined by the time-reversing operation. Since the normal form of \ldg\ is much more appropriate for calculation of the \I, there is a necessity to make a {\bf second operation of time reversal} in order to express the \coh\ \I\ by the usual \lpr.
Doing this we have to change the variables in the \ex\ for the \coh\ \I. In the language of \dg\ the \coh\ \I\ is connected with the sum of \ldg\ as follows:
\unitlength=1.0mm \lth{0.4pt}
\scalebox{0.8}{\begin{picture}(109,36)(-5,-8){
\put(11,0){\BX C\PTL\PTL
\multiput(-16.5,23)(60.5,0)2{$\Rr\+{}$}
\multiput(-16.5,-5)(60.5,0)2{$\Rr\-{}$}
\multiput(24,23)(60.5,0)2{$\Rv'\!\+\!\frac{\rv'}2$}
\multiput(24,-5)(60.5,0)2{$\Rv'\!\-\!\frac{\rv'}2$}
\Fargu{$\Rr\+1$}{$\Rr\+2$}{$\Rr\-1$}{$\Rr\-2$}
\put(-12,0)\Qq \Wlines{16,0} }
\put(49,10){\mb{\huge $=$}}
\put(72,0){\BX\LL\PTL\PTL
\Fargu{$\Rr\+1$}{$\Rr\+2$}{$\Rr\-2$}{$\Rr\-1$}
\put(-12,0)\Qq \Wlines{16,0} }
}
\end{picture}}

For the \coh\ correlation of the scattered field \fu\ it follows that:
\WBeq[GC1]
\lefteqn{\GC{\Rv,\rv}=}\NN
& &=\eot\int\!D\br{\RR_1+\rr1}D^*\br{\RR_2-\rr2}\LL(\Rab ,\rab ;\,t)
\,E\br{\Rr+2}E^*\br{\Rr-1}\,\dR_1\dR_2\dr_1\dr_2
\Eeq

Because of the singularity of the \pr\ $\LL$, the value of the integral is determined by the range of integration, for which $\Rv_1\!\approx\!\Rv_2$. Therefore we may use \FrA\ with respect to the difference $\Rv_1\-\Rv_2$:
\Beqn
D\br{\RR_1+\rr1}=
D\br{\Rv\!-\Rv_+ -\frac{\Rv_-}2\+\rr1}\!\approx D\br{\Rv\-\Rv_+}
\exp\br{\frac{ik_0}2 \Sv_{RR_+}\.(\rv\-\rv_1\-\Rv_-)},
\Eeqn
\def\FF#1#2#3{\frac{{#1}_{#2}\+{#1}_{#3}}2}
\Beqn
\text{where}\;\;\;\;\;
\Rv_+ :=\RbR 1+2\;&\;\;\text{or}&\;\;\;
\Rv_+:=(X_+,Y_+,Z_+):=\br{\FF X12,\FF Y12,\FF Z12}\\
\Rv_- :=\RaR 1\-2\;&\;\;\text{or}&\;\;\;
\Rv_-:=(X_-,Y_-,Z_-):=(X_1\-X_2,Y_1\-Y_2,Z_1\-Z_2)
\Eeqn
If we substitute the last \ex\ into the \Eq{GC1} and after integrations over $\dr_1$ and $\dr_2$ we get the \ex :
\Beq
\lefteqn{\GC{\Rv,\rv}\approx}\NN
& &\approx\!|E_0|^2\eot\! \int \!\exp\br{ik_0\Sv_{RR_+}\.\rv\,}
\!\M{D(\RR_+)}^2\Lt{\Rab}\exp\br{-\frac{Z_+}\el}
\exp\Br{-ik_0(\Sv_{RR_+}\+\Sv_Z)\.\Rv_-)}\,\dR_1\dR_2\;,
\Eeq
where $\approx$ denotes $s$-wave \app\ for $\LL$. Again converting this to its Fourier conjugated \fu\ with respect to $\rv$, we obtain:
\beq[GC2]
\GC{\Rv,\kv}=\frac\pi{2}|E_0|^2\eot\!\int\!\!\DE(\kv -k_0\Sv_{RR_+})\,\DDR+\,
\Lt{\Rab}\,\exp\br{\frac{-Z_+}\el}\exp\!\Br{ik_0(\Sv_{RR_+}\!\+\Sv_Z)\.\Rv_-}
\dR_1\dR_2\;,
\eeq
where:
\Beq
\DELTA+
\Eeq
As in the case of \ic\ \sc\ eqs. \Eq{JJL}-\Eq{GL} we can obtain the \I\ in the point $R\!=\!(0,0,Z)$ for a given direction $\Sv\!=\!(\SIA,0,\COA)$:
\Beq[GC]
\lefteqn{\JCZ\frac{\pi\eot}\MC \Int_{\! Z_+=0}\!\!\TZ+\EZP
\int_{\! Z_-=-2Z+}^{+2Z+}\!\exp\Br{-ik_0\,(1\!\+\COA)Z_-}
\times }\NN
& &\times\int_{\!\rho\in{\Bbb R}^2}\!\e^{-i\kp\!\.\rov}\!
\Br{\LL (\sqrt{\rho^2+Z_-^2}\,,\: t)-\LL (\SQ{2Z_+\+2Z_0}\,,\: t)}\;
d^2\rov\,\dZ_-\dZ_+\;.
\Eeq

Similarly to eq. \Eq{FL}, we can write:
\beq[FC]
\JC 8\NORM\,\Fg [\HC{u}],
\eeq
where $H^C$ is defined by
\Beq
\lefteqn{\HC{u}:=\frac{1}{2\MC}\,\Int_{a=0}\!\Theta\br{\frac{h-a}{\COA}}
\exp\br{\frac{a-h}{\COA}-a}\int_{b=-2a}^{2a}\!\!\exp\Br{-ik_0\el\,(1\!\+\COA)b}
}\NN
& &\Int_{\rho=0}\e^{-i\kp\!\.\rov}\!
\Br{\EXF{\sqrt{\rho^2\+b^2}} -\EXF{\SQ{2a\+2M}} }
\:\rho\dro\,\db\da
\Eeq
If we introduce the notations $\xi(u):=2\kls{u^2} \,,\;\;\;\;\eta:=2k_0\el(1\!+\!\COA)$ we get the following \ex\ for $H^C$:
$$\!\HC u =
\frac{1}{\xi^2+\eta^2} \left\{ \frac{\e^{-h}\-\TECA}{1-\COA }
-\frac{\EaH [\AXC \CEH +\eta\COA\SEH]-\TECA}{\ZNA}+\right.$$
\Beq[HC]
\left.+\Br{\frac{\xi^2\+\eta^2}{\xi }(1\-\e^{-M\xi})-\xi}
\frac{\EaH [\AXC\frac{\SEH}{\eta}-\COA\CEH]+\COA\,\TECA}{\ZNA}
\right\}
\Eeq
Finally for the \coh\ \MSC\ \I\ we obtain:
\beq[JC]
\JC 8\NORM\Br{\Dterm{\GT}{\HC{\xt{}}}+\Iterm{\GT}{\HC u}}
\eeq
The connection between $\chi$ and $g$ is defined by eq.\Eq{EDL}. Following eq.\Eq{FC} and eq.\Eq{JC} the \coh\ \I\, for the \st\ case without absorption is obtained :
\beq[JCS]
\JC 8\NORM \FU1{\HC u}= 8\NORM \Br{3\,\HC0 +\Iterm1{\HC u}}
\eeq

The summary \sc\ \I\ is represented on Fig.~\ref{JLC} and the single \sc\ \I\ on Fig.~\ref{JL1}.
\bfig
\unitlength=1.0mm \lth{0.4pt}
\begin{picture}(170,128)(13,3)
\epsfig{file=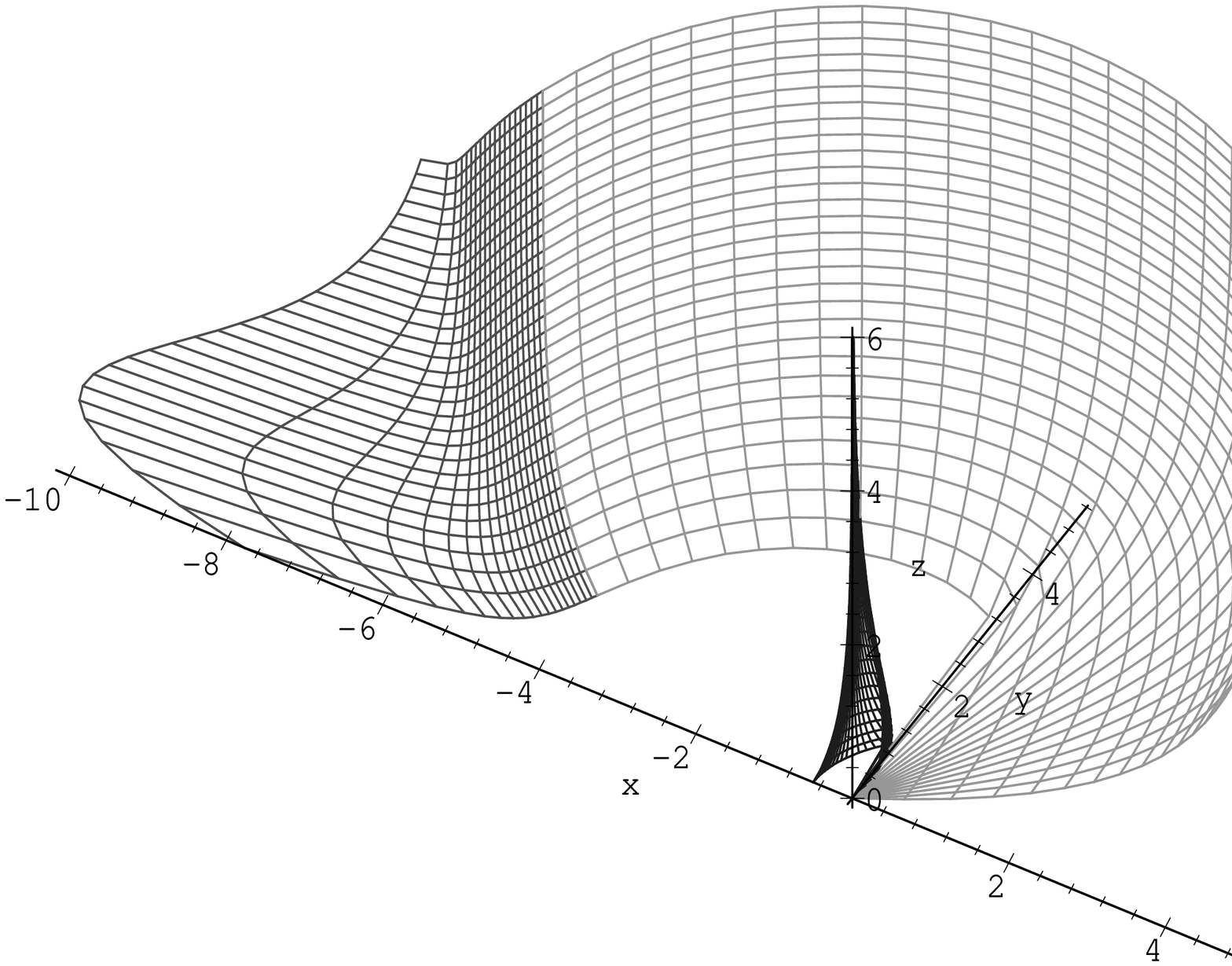, width=6.5in,height=8.2in}
\end{picture}
\caption{Multiple \sc\ \I.}
\label{JLC}
\vspace{2em}
\begin{picture}(170,90)(-15,5)
\epsfig{file=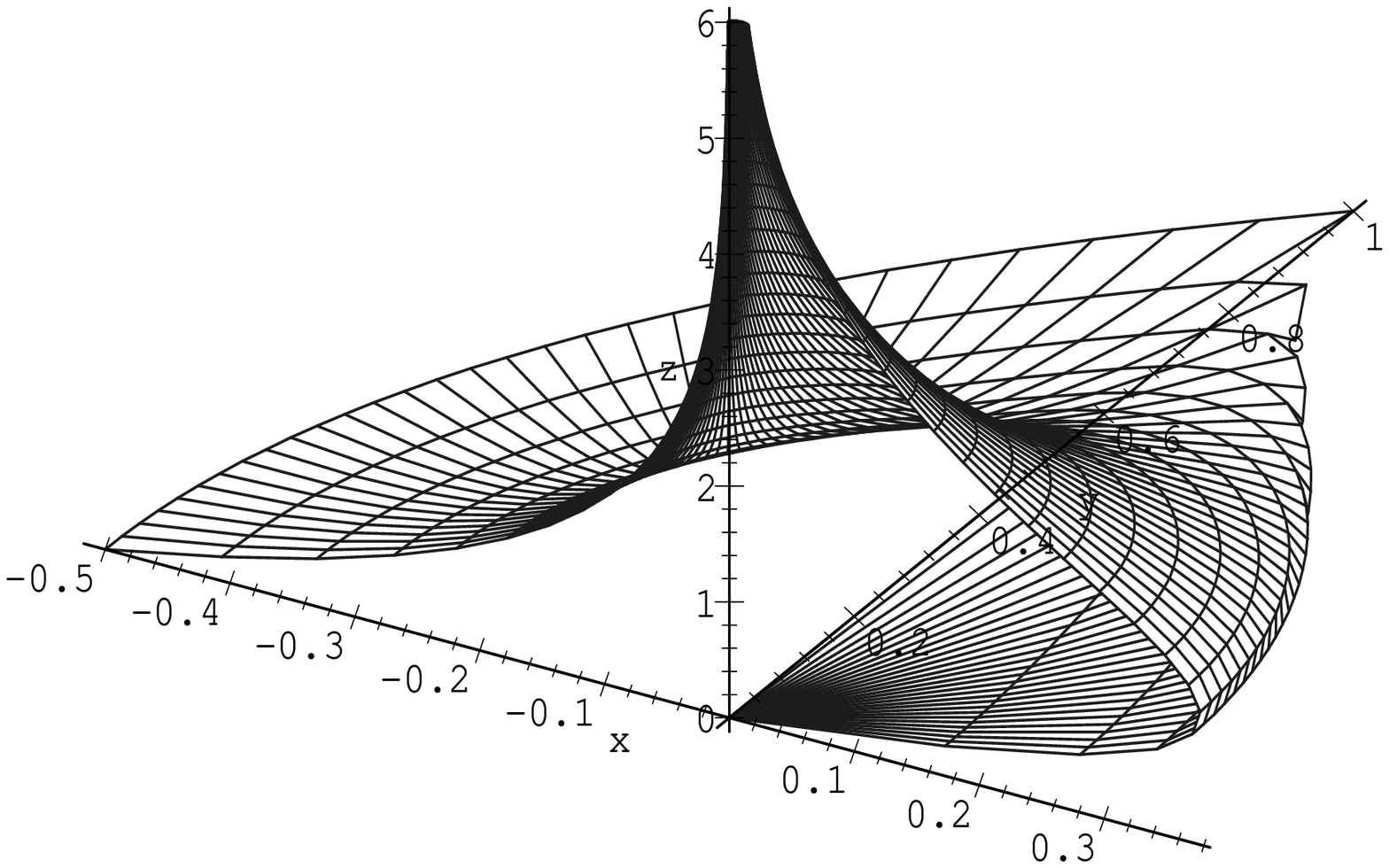, width=5.5in, height=7.5in}
\end{picture}
\caption{Single \sc\ \I.}
\label{JL1}
\efig
\end{widetext}
In these cylindrical plots the polar variable corresponds to the \sc\ angle, the radial variable corresponds to the \sc\ \I, and the vertical variable is the depth of observation in dimensionless units (\EMFP). $k\el=100$ Forward direction is in the right side of the plots.

\section{Partial Intensities.}
In this section we shall estimate the validity of the \DA\ in different orders of \MSC. Our goal is to find a physically well-motivated explanation for the failure of the diffusion \app\ in the case of large \sc\ angles.

We have shown (eqs.\Eq{GS},\Eq{Kn} and \Eq{LK}) that the \FI\ of the \lpr\ are a convergent infinite geometric series. It is therefore possible to study the scattered \I\ as a \fu\ of the order of \sc. It means that the \coh\ and \ic\ intensities may be expanded into the orders of \MSC:

$J^{\LL}\!=\!\sum_{n=2}^\infty J_n^L\,;\;J^C\!=\!\sum_{n=2}^\infty J_n^C$, where $J^L_n$ and $J^C_n$ denote, respectively, the partial \ic\ and \coh\ intensities of $n^{\text{th}}$ order.

To obtain the partial intensities, we start from the \ex s \Eq{GL} and \Eq{GC}, in which $\Lt{\rov,Z}$ is represented by its \FI s:
\def\TECO{ {\bf \theta}\br{\,\frac\pi2 \-\alpha}\exp\br{-\frac{h}\COA} }
\def\qeta{q\-\eta}
\WBeq[p1]
\lefteqn{J^{\LL} (Z,\alpha,t)=
\frac\Eot{4\MC}\,
\Int_{Z_1=0}\TZ1\,\EZ1\Int_{\! Z_2=0}\!\e^{-\frac{Z_2}\el}
\int_{\rov\in{\Bbb R}^2}\!\!\e^{-i\kp\.\rov}\,\times
}\NN
& & \Int_{K_z=-\infty}\int_{\Kp\!\in {\Bbb R}^2 }\!
\e^{i\Kp\.\rov}
\Br{\e^{iK_z\,(Z_1\- Z_2)}-\e^{iK_z\,(Z_1\+Z_2\!+2Z_0)}}\,
\LL\br{\sqrt{K_\perp^2\!\+K_z^2},\,t}\:
\frac{\d^2\Kp}{(2\pi)^2}\,\d K_z\,
\d^2\rov\:\dZ_2\dZ_1
\Eeq
\ \\
\Beq[p2]
J^C(Z,\alpha,t)=
\frac\Eot{4\MC}\,
\Int_{Z_+=0}\TZ+\,\EZP\,
\int_{Z_-=-2Z_+}^{2Z_+}\!\exp\Br{-ik_0\,(1\!\+\COA)Z_-}\,\times  \NN
\int_{\rov\in{\Bbb R}^2}\!\!\e^{-i\kp\!\varepsilon\.\rov}\!
\Int_{K_z=-\infty}\int_{\Kp\!\in {\Bbb R}^2 }\!\e^{i\Kp\.\rov}
\Br{\e^{iK_z\,Z_-}-\e^{i2K_z\,(Z_+\!+Z_0)}}\,
\LL\br{\sqrt{K_\perp^2\!\+K_z^2},\,t}\:
\frac{\d^2\Kp}{(2\pi)^2}\,\d K_z\,
\d^2\rov\:\dZ_-\dZ_+ \;,
\Eeq
where $|\kp|\!:=\!k_0\SIA$. Performing integrations, first with respect to $\d^2\rov$ and $\d^2\Kp$ and after that with respect to $\d Z_2\,\d Z_1$ in \Eq{p1}, and $\d Z_-\,\d Z_+$ in \Eq{p2}, it follows:

\beq[p3]
J^{\LL}\hat\frac\Eot4\!\!\InT\!\el\,\LL\br{\frac{p}\el,\,t}
\frac{\e^{ihp}-\TECO}{1+ip\COA}\br{\frac1{1+ip}-
\frac{\e^{2iMp}}{1-ip}}\d p
\eeq
\ \\
\Beq[p4]
\lefteqn{J^C\hat }\NN
& &=\frac{|E_0|^2\e^{-i\o0 t -h}}{2\MC}\!\!
\InT\!\el\,\LL\br{\frac{\sqrt{q^2\+v^2}}{2\el},\,t}\!\!\left[
\frac{ \cos((\qeta)h)\+\frac\beta{q-\eta}\sin((\qeta)h)}{(\qeta)^2+\beta^2}-
\e^{i(h+M)q}\frac{\cos(\eta h)\+
\frac{(\beta-iq)}\eta\sin(\eta h)}{(\beta\-iq)^2+\eta^2}
\right]\!\!\d q\;, \NN
& &\text{where\ }\;\;\;\;\; h:=\frac Z\el\;,\;\;p:=K_z\el\;,\;\;q:=2p\;,
\;\;v:=2k_0\el\SIA\;,\;\;
\beta:=1\+\frac1\COA\;,\;\;\eta:=2k_0\el(1\+\COA)\;.
\Eeq

As we already mentioned the exact time-dependent \lpr\ is geometric series with base $\gt\Atg p$:

\beq
\LL\br{\frac{p}l,\,t\!}=\frac{4\pi}\el \frac{g^2(t)\Atg p}{1-\gt\Atg p}=
\frac{4\pi}\el \gt\sum_{n=1}^\infty \br{\gt\Atg p}^{\!n}
\Weeq

To compare the exact partial intensities with these, corresponding to the \DA, we should find the correct \rep\ of the diffusion \pr\ as convergent geometric series. In the standard approach \cite{Gb,McKSJ,McKSJ2,St,AWM}, the diffusion \pr\ is considered as obtained by the exact \lpr\ taking only the first order with respect to $p^2$ for $|p|\!\ll\!1\,$ in the expansion of $\arctg (p) /p$, so that:
\Beq
\gt\Atg p\,\rightarrow\, 1-\frac13(\chi^2_1(t)\+p^2)
\Eeq
and
\Beq[p6]\nonumber
\lefteqn{\frac\gt{1-\gt\Atg p}\,\rightarrow}\\
&\rightarrow\,\frac{3\,\gt}{\Xt1+p^2}
\stackrel{?}{=}
3\,\gt\sum_{n=0}^{\infty}\br{\,1-\Xt1\-p^2\,}^n,\NN
&\text{for\ }\; |p|<\sqrt{1\-\Xt1}\;,\;\;\;\chi_1(t)\ll1\;.
\Eeq

Thus it is not clear in which way should be performed the \Ftr\ over $p$ because of the restriction to the $p$-values ($|p|\!\ll\!1$). The above mentioned \app\ is used in paper \cite{AWM} in order to be obtained the partial \coh\ intensities in the \st\ case without absorption ($\chi\!=\!0$). To obtain the different orders of \sc\ one must take the \FI\ of the different addends of $\sum_{n=0}^{\infty}(1\-p^2)^n$. Since the convergence radius of the sum in \Eq{p6}, imposes a restriction to the $p$-values ($|p|\!<\!1$), we face with the cut-off of the integral limits: $\InT\d p\!\rightarrow\!\int_{|p|<1}\d p\;.$ Therefore, the partial \coh\ intensities turn out to be zero outside the interval $0\!\leq\!\pi\-\alpha\!<\!1/k\el$. In contrast, the total \coh\ \I, which has been calculated after integrating over all real $p$, does not become zero outside the polar angle interval. So that the summation of all \sc\ orders can not reproduce the total \coh\ \I. In paper \cite{AWM} this cut-off of the partial intensities is interpreted as a feature of the \DA, but in our opinion it is {\bf an additional systematical error itself, which does not originate from the \DA}. This strange situation can be escaped by a rather different choice of the geometric series base, standing on the following considerations:

1. {\bf The single-\sc\ has not to be included in the diffusion \pr\ }
\cite{Gb}.

2. The \DA\ for \lpr\ describes the asymptotic behaviour of $\Lt R$ for large values of $R$ and corresponds simply to the first term of the \ex \Eq{LRt}. {\bf So the diffusion \pr\ in conjugated variables must be considered as a \FI\ of the first term of \Eq{LRt}, but not as an \app\ for the \FI\ of the exact \ex\ \Eq{p6}, when $|p|\!\ll\!1$}. Namely \Eq{LRt} could be written as:
\def\Bsr{\frac B{B+p^2\!+\!\Xt{}}}%
\def\ftrp{e^{i\pv\,\.\frac\Rv\ell}}
\begin{widetext}
$$\Lt R=\frac{4\pi}{\el^4}\!\int\!\!\frac{g^2(t)\Atg p}{1-\gt\Atg p}
\,\ftrp\,\frac{\D\pv}{(2\pi)^3}=\frac{4\pi}{\el^4}\!
\int\!\!\frac{A(\xt{})}{p^2\+\Xt{}}\,\ftrp\,\frac{\D\pv}{(2\pi)^3}\,+\,
\frac1\Rl\!\Iterm\GT{\E R}$$
\end{widetext}
3. {\bf The diffusion \pr\ in $p$-\rep\ must be an infinite geometric sum with series base between 0 and 1 not only for small $p$-values, but also for all real $p$-values.} This allows us to take \FI\ of $\L_n(p/\el,t)$ without imposing restrictions to the integral limits. In the context of above mentioned arguments, the exchange $\gt\Atg p\rightarrow\Bsr$ gives the most proper choice of series base. In contrast to the standard approach the last one is not an \app\ for the exact series base. Then
\Beq[p7]
\frac{g^2(t)\Atg p}{1-\gt\Atg p}=\gt\sum_{n=1}^\infty \br{\gt\Atg p}^{\!n}
\;\;
\stackrel{{\rm diff}}{\longrightarrow}\NN
\stackrel{{\rm diff}}{\longrightarrow}
\;\;\frac A{p^2\+\Xt{}}=
\frac AB\sum_{n=1}^{\infty}\br{\Bsr}^n\;,\NN
\Eeq
where $A$ has already been defined in \Eq{LRt} and $B$ is an arbitrary positive \fu, moreover it may depend on $p$.

4. To restrict the generality of $B$ ($B\!:=\!B(p,\chi)$), according to the correspondence between the two sums in \Eq{p7}, it should be determined $B$ as $B\!:=\!A(\xt{})\GT$. Besides $\gt\Atg p$ and $\Bsr$ have one and the same behaviour for $p$-values close to the complex singularities $\pm i\chi$ of the \lpr :
$$\gt\Atg p=1-\frac\gt{A}(p^2\+\chi^2)\,+
\;\;\text{higher orders\ }\; p^2\+\chi^2\;,$$
$$\Bsr=1-\frac1B (p^2\+\chi^2)\,+\;\;\text{higher orders\ }\; p^2\+\chi^2\;.$$
Furthermore, both series bases vanish for infinite $|p|$.

The above considerations lead us to the choice \Eq{p7}, which avoids any problems with the \Ftr. Moreover, we find decomposition of the diffusion \pr\ to the different orders of \sc\ not only for large EDL (small $\chi$), where $\chi$ is given by its first order approximated \rep\ $\xt1:=\sqrt{3\GT\-3}$, but also for any $\chi$, which is a solution of the \eq\ \Eq{EDL}. Further we are going to demonstrate that the sum of the partial intensities tends really to the total scattered \I\ when the number of included \sc\ orders increases significantly. In addition, for fixed number of included \sc s, the sum approaches to the total \I\ when the EDL increases.

\subsection{Incoherent intensities.}
Replacing the \lpr\ in \Eq{p3} with $L_{n+1}(p;t)\!:=\!(4\pi\gt/\el)\,(\gt\,\arctg\,p/p)^n$, we will get the $(n\+1)$-th partial \I:
\Wbeq[p8]
J_{n+1}^L\hat\pi\Eot\gt\!\InT\!\br{\gt\Atg p}^{\!n}
\frac{\e^{ihp}-\TECO}{1+ip\COA}\br{\frac1{1+ip}-\frac{\e^{2iMp}}{1-ip}}\d p\;,
\eeq
and after the exchange $\Lt{p}\rightarrow\L_{n+1}^{\rm diff}(p,t)\!:=(4\pi\gt/\el)\,(B/(B\+p^2\+\Xt{}))^n$ in \Eq{p8}, we will have the partial intensities in \DA:
\beq[p8a]
J_{n+1}^{L^{\rm diff}}\hat\pi\Eot\gt\!\InT\!\br\Bsr^{\!n}
\frac{\e^{ihp}-\TECO}{1+ip\COA}\br{\frac1{1+ip}-\frac{\e^{2i\Mas p}}{1-ip}}
\d p\;,
\eeq
Upper \ex\ can be evaluate analytically as an contour integral in complex plane, using the theorem of residuums, so we get:
\def\COSU#1{\frac{\br{\frac\COA{u\COA-1}}^{k-m}\!\!-
\br{\frac1{u{#1}1}}^{k-m}}{1{#1}\COA} }
\Beq[p9]
\lefteqn{
J_{n+1}^{L^{\rm diff}}\hat 2\NORM\gt
\left\{\; \br{\frac B{u^2\-1}}^{\!n}\frac{\e^{-h}\-\TECO}{1-\COA}\,+
\hspace{50pt}u:=\sqrt{B\+\Xt1}
\right.}\NN
& &+\,B^n\!\sum_{k=1}^n{2n\-k\-1 \choose n\-1}(2u)^{k-2n}
\sum_{m=0}^{k-1}\frac1{m!}
\left[\COSU- \br{h^m\e^{-uh}-\theta(-m)\TECO}+\right.
\NN
& &\left.\left.+\,\COSU+ \br{(h\+2\Mas)^m\e^{-u(h+2\Mas)}-(2\Mas)^m\e^{-2\Mas u}\TECO }
\right]\right\}\;.
\WEeq

For first order \app\ \ $\xt{}\!\approx\!\xt1\!:=\!\sqrt{3\GT\-3}\:$, $\:A(\xt{})\!\approx\!A_1(\xt{})\!:=\!3\:$ are given in appendix, for the angles $\:\alpha\!=\!\{0 , \pi\!/2 , \pi\}\:$ consequently \Eq{p20}, \Eq{p21}, \Eq{p22}.

\bfigW
\epsfig{file=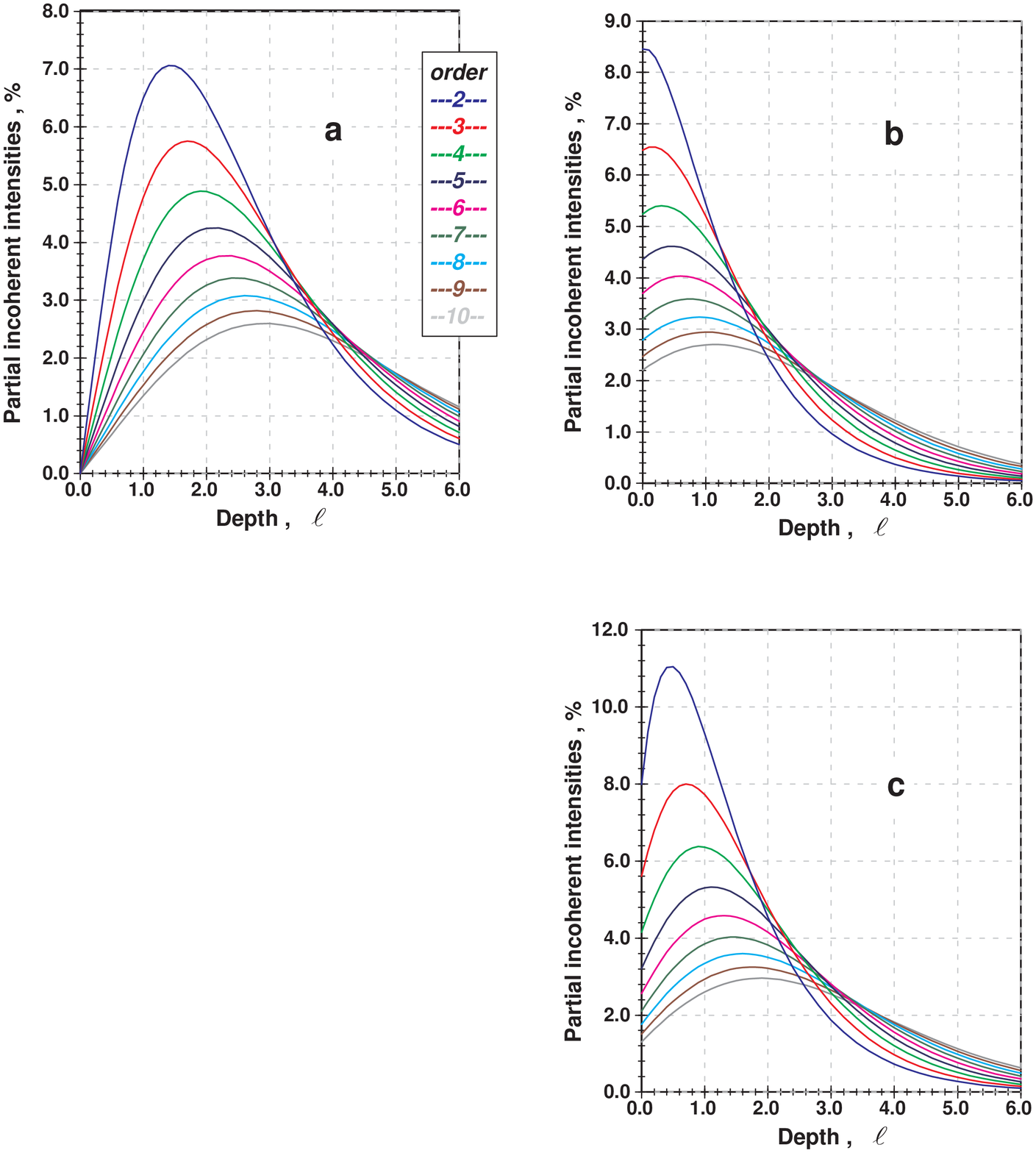, width=6.8in}
\put(-175,50){
\begin{tabular}{|l|c|c|c|}
\hline
\vspace{-0.7ex}
&  &  &  \\
$\;\pi\!-\!\alpha$ &$\sum_{n=2}^{30} J^C_n(\alpha)$
&$J^C(\alpha)$&$\sum_{n=2}^{30} J^C_n(\alpha)/J^C(\alpha)$\\
\vspace{-0.7ex}
&  &  &  \\
\ \ \ [mrad]&\ \ \ \ [\%] &\ \ \ [\%]&\ \ \ \ \ [\%] \\
\vspace{-0.7ex}
&  &  &  \\
\hline
\vspace{-0.9ex}
&  &  &  \\
{\bf\ 0}&{\bf 60.1602}&{\bf 100}&{\bf 60.1602}\\
\ 0.1 & 58.6109 & 79.3561 & 73.8581\\
\ 0.2 & 54.4049 & 64.1648 & 84.7894\\
\ 0.3 & 48.6108 & 52.7181 & 92.2089\\
{\bf\ 0.328259}&{\bf 46.8503}&{\bf 50}&{\bf 93.7007}\\
\ 0.4 & 42.3775 & 43.9175 & 96.4935\\
\ 0.5 & 36.5173 & 37.0323 & 98.6094\\
\ 0.6 & 31.4086 & 31.5630 & 99.5109\\
\ 0.7 & 27.1179 & 27.1597 & 99.8459\\
\ 0.8 & 23.5616 & 23.5719 & 99.9560\\
\ 0.9 & 20.6147 & 20.6171 & 99.9885\\
\vspace{-0.9ex}
{\bf\ 1.0}&{\bf 18.1591}&{\bf 18.1596}&{\bf 99.9972}\\
&  &  &  \\
\hline
\end{tabular}
}
\caption{Partial \ic\ intensities for various depth for forward (\ref{L_par}a), backward (\ref{L_par}b) and perpendicular (\ref{L_par}c) directions (100\% corresponds to total \ic\ \I\ on the surface).}
\label{L_par}
\efigW

From 2 to 10 order \ic\ \sc\ \I\ for forward (Fig.~\ref{L_par}a), backward (Fig.~\ref{L_par}b) and perpendicular (Fig.~\ref{L_par}c) direction are graphically represented, as functions of depth calculated by formula \Eq{p9} for stationary case $B\!=\!3$. There is a small difference between the results coming from \Eq{p8} and these one's given by \Eq{p9}, which becomes negligible for depths of order several \MFP.

\subsection{Coherent intensities.}
Replacing the \lpr\ in \Eq{p4} with
$L_{n+1}(\tilde{p};t)\!:=\!(4\pi\gt/\el)\,(\gt\,\arctg\,\tilde{p}/\tilde{p})^n$, where $\tilde{p}:=\sqrt{q^2\+v^2}/2$,
we will get the $(n\+1)$-th partial \I:
\WBeq[p10]
\lefteqn{J_{n+1}^C\hat \frac{2\pi\,\gt\,|E_0|^2\,\e^{-i\o0 t -h}}\MC\,\times}\NN
& &
\times\,\InT\!\br{\gt\frac{\arctg\br{\sqrt{q^2\+v^2}/2}}{\sqrt{q^2+v^2}/2}}^{\!\!n}\!\left[
\frac{ \cos((\qeta)h)\+\frac\beta{q-\eta}\sin((\qeta)h)}{(\qeta)^2+\beta^2}-
\e^{i(h+M)q}\frac{\cos(\eta h)\+
\frac{(\beta-iq)}\eta\sin(\eta h)}{(\beta\-iq)^2+\eta^2}
\right]\!\!\d q\;,
\Eeq

Following \Eq{p7} it can be written:
\Beq[p11]
\lefteqn{
J_{n+1}^{C^{\rm diff}}(h,\alpha,t)=
\frac{2\pi\,\gt\,|E_0|^2\,\e^{-i\o0 t -h}}\MC\,\times
}\NN
& &
\times\,\InT\!\br{\frac{4B}{4(B\+\chi^2)\+q^2\+v^2}}^{\!n}
\!\!\Br{
\frac{ \cos((\qeta)h)\+\frac\beta{q-\eta}\sin((\qeta)h)}{(\qeta)^2+\beta^2}-
\e^{i(h+\Mas)q}\frac{\cos(\eta h)\+
(\beta\-iq)\frac{\sin(\eta h)}\eta}{(\beta\-iq)^2+\eta^2}
}\!\!\d q
\Eeq
and for stationary case ($\xt{}\!=\!0,\;B\!=\!3$)
\Beq[p12]
\lefteqn{
J_{n+1}^{C^{\rm diff}}(h,\alpha,t)=
\frac{2\pi\,\gt\,|E_0|^2\,\e^{-i\o0 t -h}}\MC\,\times
}\NN
& &
\times\,\InT\!
\br{\frac{12}{q^2\+v^2\+12}}^{\!n}
\!\!\Br{
\frac{ \cos((\qeta)h)\+\frac\beta{q-\eta}\sin((\qeta)h)}{(\qeta)^2+\beta^2}-
\e^{i(h+\Mas)q}\frac{\cos(\eta h)\+
(\beta\-iq)\frac{\sin(\eta h)}\eta}{(\beta\-iq)^2+\eta^2}
}\!\!\d q
\Eeq

\Beq
\lefteqn{
J_{n+1}^{C^{\rm diff}}(h,\alpha,t)=
\frac{2\pi^2\,\gt\,|E_0|^2\,\e^{-i\o0 t -h}}{\MC\beta}
\left\{\br{\!\frac{4B}{(\eta\+i\beta)^2+w^2}\!}^{\!\!n}e^{\beta h}+
\br{\!\frac{B}{w^2}\!}^{\!\!n}\sum_{k\!=\!0}^{n\-1} (2w)^{k\+1}{2n\-2\-k \choose n\-1}\times
\right.}\NN
& &\!\!\times\!\left[(\-1)^k\!\!\sum_{m=0}^k\frac{h^m}{m!} e^{(w\+i\eta)h}\!
\Br{(w\+i\eta)^{m-k-1}\!-(w\-\beta\+i\eta)^{m-k-1}}-
(\-1)^m e^{-(w\+i\eta)h}\!
\Br{(w\+i\eta)^{m-k-1}\!-(w\+\beta\+i\eta)^{m-k-1}}\right. \NN
& &\left.\left.-e^{-(h+M)w}\frac\beta{i\eta}\!\sum_{m=0}^k\!\frac{(h\+M)^m}{m!}
\Br{\frac{e^{i\eta h}}{(\beta\+w\-i\eta)^{k-m+1}}-\frac{e^{-i\eta h}}{(\beta\+w\+i\eta)^{k-m+1}}}
\right]\right\}
\WEeq

\begin{figure}\label{Cpar}
\epsfig{file=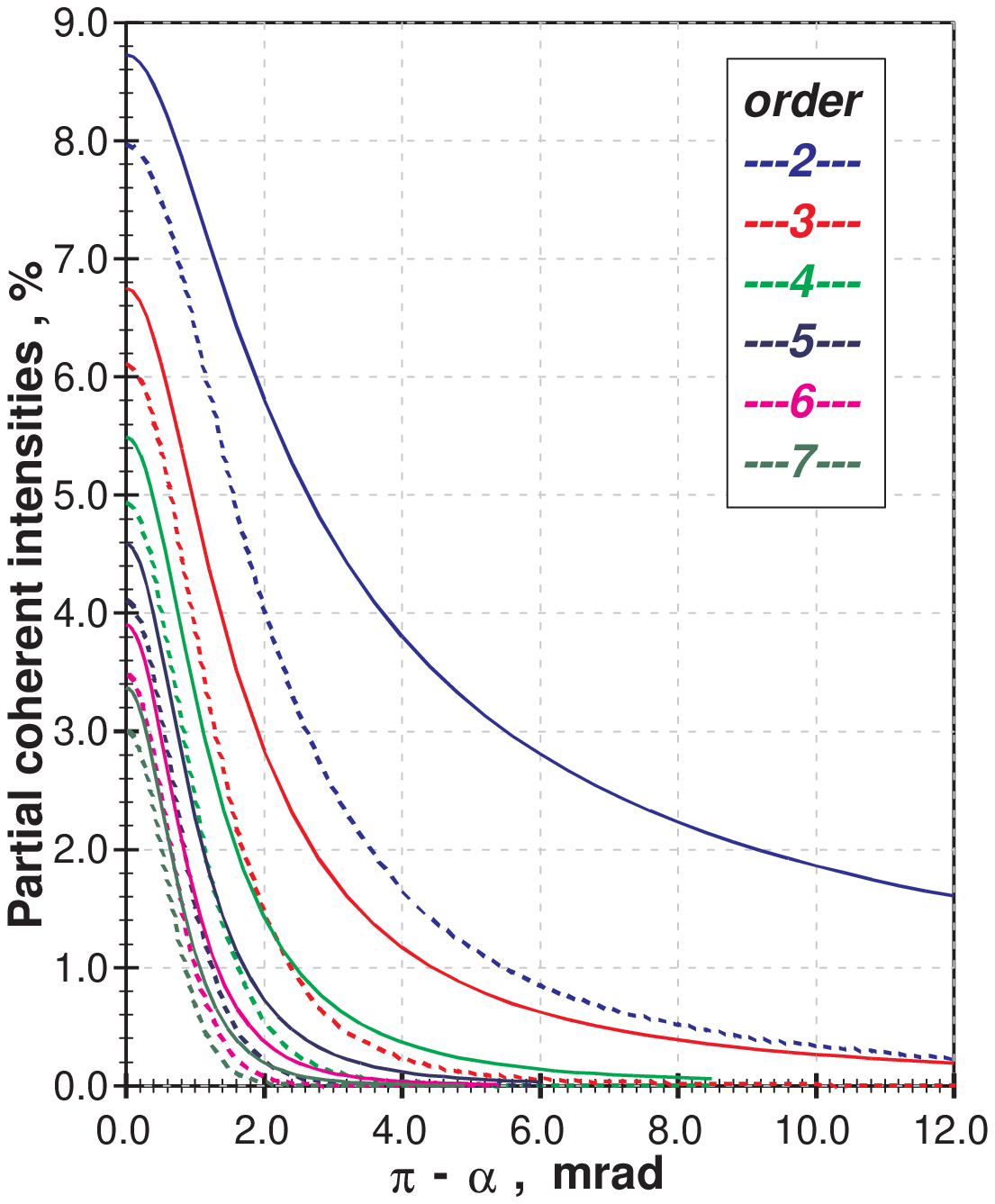, width=3.4in}
\caption{The exact lineshapes represented by solid lines and the lineshapes for \DA, represented by dashed lines (100\% corresponds to total \coh\ \I\ on the surface).}
\end{figure}

The exact \ex s \Eq{p10} give us an opportunity to take into consideration the contributions of the different orders of \MSC. It afford us also illustrate the limits of applicability of the \DA\ for the different orders of \sc\ Fig.~\ref{Cpar}.

\ \\
\noindent\begin{tabular}{|c||c|c|c|c|c|}
\hline
&\multicolumn{4}{c}{}&\\
$\;\pi\!-\!\alpha $&\multicolumn{4}{c}{$\sum_{n=2}^{10} J^C_n(\alpha)/J^C(\alpha)$\ [\%]}&\\
{\small [mrad]} &\multicolumn{4}{c}{}&\\
\cline{2-6}
\vspace{-1em}
& & & & &\\
&$\chi=0$&$\chi=0.01$&$\chi=0.1$&$\chi=0.2$&$\chi=0.5$\\
\vspace{-1em}
& & & & &\\
\hline
\vspace{-0.9em}
& & & & &\\
{\bf\ 0.0}&{\bf 38.8935}&{\bf 39.8374}&{\bf 48.6249}&{\bf 58.7508}&{\bf 86.4635}\\
\ 0.1 & 48.4693 & 48.5190 & 52.6263 & 60.9547 & 86.9847 \\
\ 0.2 & 58.0091 & 58.0342 & 60.3816 & 66.2827 & 88.4047 \\
\ 0.3 & 66.9634 & 66.9796 & 68.5276 & 72.7159 & 90.3713 \\
\ 0.4 & 74.8999 & 74.9110 & 75.9879 & 78.9887 & 92.4898 \\
\ 0.5 & 81.5625 & 81.5704 & 82.3306 & 84.4765 & 94.4541 \\
\ 0.6 & 86.8802 & 86.8857 & 87.4206 & 88.9389 & 96.0933 \\
\ 0.7 & 90.9328 & 90.9366 & 91.3085 & 92.3661 & 97.3566 \\
\ 0.8 & 93.8955 & 93.8981 & 94.1527 & 94.8765 & 98.2711 \\
\ 0.9 & 95.9834 & 95.9852 & 96.1565 & 96.6431 & 98.9007 \\
\vspace{-0.9em}
{\bf\ 1.0}&{\bf 97.4086}&{\bf 97.4097}&{\bf 97.5232}&{\bf 97.8449}&{\bf 99.3168}\\
& & & & &\\
\hline
\end{tabular}
\begin{widetext}\center
\begin{tabular}{|c||c|c|c|c|c|c|c|c|c|}
\hline
$\pi-\alpha, mrad$ &2 order&3 order&4 order&5 order&6 order&7 order&8 order&9 order&10 order\\
\hline
& & & & & & & & &\\
0&\ 0.91352 \ &\ 0.90569 \ &\ 0.89992 \ &\ 0.89563 \ &\ 0.89347 \ &\ 0.89323 \ &\ 0.89444 \ &\ 0.89681 \ &\ 0.90625 \ \\
0.6 & 0.89083 & 0.86237 & 0.83804 & 0.81649 & 0.79813 & 0.78263 & 0.76946 & 0.75833 & 0.75783\\
1.2 & 0.82067 & 0.73078 & 0.65418 & 0.58712 & 0.52865 & 0.47765 & 0.43297 & 0.39394 & 0.37374\\
1.8 & 0.72441 & 0.56627 & 0.44525 & 0.35067 & 0.27672 & 0.21888 & 0.17357 & 0.13834 & 0.12712\\
2.4 & 0.62918 & 0.4242  & 0.28784 & 0.19555 & 0.13297 & 0.09052 & 0.06172 & 0.04243 & 0.04917\\
3.0 & 0.54681 & 0.31814 & 0.18639 & 0.10935 & 0.06418 & 0.03769 & 0.02215 & 0.01323 & -0.05875\\
12 & 0.1598  & 0.02602 & 0.00425 & 0.00069 & 0.00007 & 0.00001 & 5.28693E-7 & -3.68948E-8 & -2.42003E-11\\
& & & & & & & & &\\
\hline
\end{tabular}
\end{widetext}

\section{Back to the diffusion approximation and small angle analysis.}
It is easy to obtain the results for \ic\ and \CSC\ intensities in the \DA\ from eqs. \Eq{JL} and \Eq{JC}. It is sufficient to take the first terms in eqs. \Eq{JL} and \Eq{JC} and after that to take into account the \app\ $N\!\!=\!\!1$ eq. \Eq{o1} in the \DT, so $\chi\!\!\approx\!\!\chi_1$ and the \ic\ \I\ becomes:
\beq
\JL 6\NORM H^{\LL}(\chi_1(t),h,\alpha)\ .
\eeq
For \coh\ \I\ the result is:
\beq
\JC 24\NORM H^C(\chi_1(t),h,\alpha)\ .
\eeq
As it is seen from the graphics (Fig.~\ref{Cpar}), the \DA\ for the \CBSC\ gives a good description only in a narrow cone around $\alpha\!\approx\!\pi$.
In the region of the \BSC\ peak \ic\ \I\ is practically a constant. In most previous papers this has been understood without mentioning explicitly:
\beq
\HL{\chi_1(t)}\!\approx\! H^{\LL}(\chi_1 (t),h,\pi)
\eeq
Using the \ex\ \Eq{HL} and \Eq{JC} for \ic\ \I\ in depth it can be seen that
\Wbeq[Lda]
\JL 3\NORM\Br{\frac{\e^{-\chi_1(t)h}}{(1\+\chi_1(t))^2}\br{1\+
\frac{1\-\e^{-2M\chi_1(t)}}{\chi_1(t)}}+\e^{-\chi_1(t)h}\+\e^{-h}}
\Weeq
The simple approximated formulas for \WL\ term can be obtained after substitution $\eta\!=\!0$ in \Eq{HC}. In the standard algebras(calculations) of the \coh\ \I\ \ex , such an assumption is made before integration over $\dR$ \cite{McKSJ}.
\WBeqn
J^C(h,\alpha\!\approx\!\pi,t)\approx\frac{24\NORM}{\xi^2}
\BR{\frac{\e^{-h}}2 -\frac\EaH{2\+\xi}
\Br{\xi\e^{-M\xi}\br{h+\frac1{2\+\xi}}\+1} }
\WEeqn
Now let's take $h\!=\!0$ and we obtain the well known \ex s for the \ic and \CSC\ intensities on the surface.
\beq[JL2]
J^{\LL}(0,\alpha,t)\approx\frac{3\NORM}{(1+\chi_1(t))^2}(1\+ 2M)
\eeq
\Wbeq[JC2]
J^C(0,\alpha,t)\approx\frac{3\NORM}{\br{1+\kls{\Xt1}\;}^2}
\Br{1+\frac{1\-\exp\br{-2M\kls{\Xt1}\;}}{\kls{\Xt1}}}
\Weeq

\section{Integral scattering intensities for various depths in the stationary case.}

In some cases like these devoted to the investigations of the relationship between the impurities averaged perturbation theory and classical transport theory \cite{RKT,St1,Cl}, the integral \I\ over all directions must be included. However these considerations face some difficulties when one tries to take into account the \WL\ effects. We have not intention to deal with this problem now. Independently it is interesting to know the difference between the integral \sc\ \I\ in the \DA\ and this one determined by the \ex s \Eq{JLS}, \Eq{HL}, \Eq{JCS} and \Eq{HC}, so let's introduce the ratios:

\beq[RL]
R_L(h):= \frac{3\, \int_0^{\pi} \HL0\SIA\d\alpha}
{\int_0^{\pi} \Br{3\,\HL0 +\Iterm1{\HL{u}}}\SIA\d\alpha}
\eeq

\beq[RC]
R_C(h):= \frac{3\,\int_0^{\alpha_0}\HC0\SIA\d\alpha}
{\int_0^{\alpha_0}\Br{3\,\HC0 +\Iterm1{\HC{u}}}\SIA\d\alpha}
\eeq

\def\Fileh#1{\center \epsfig{file={#1}.eps, width=5in, height=2in}}

\begin{figure}
\epsfig{file=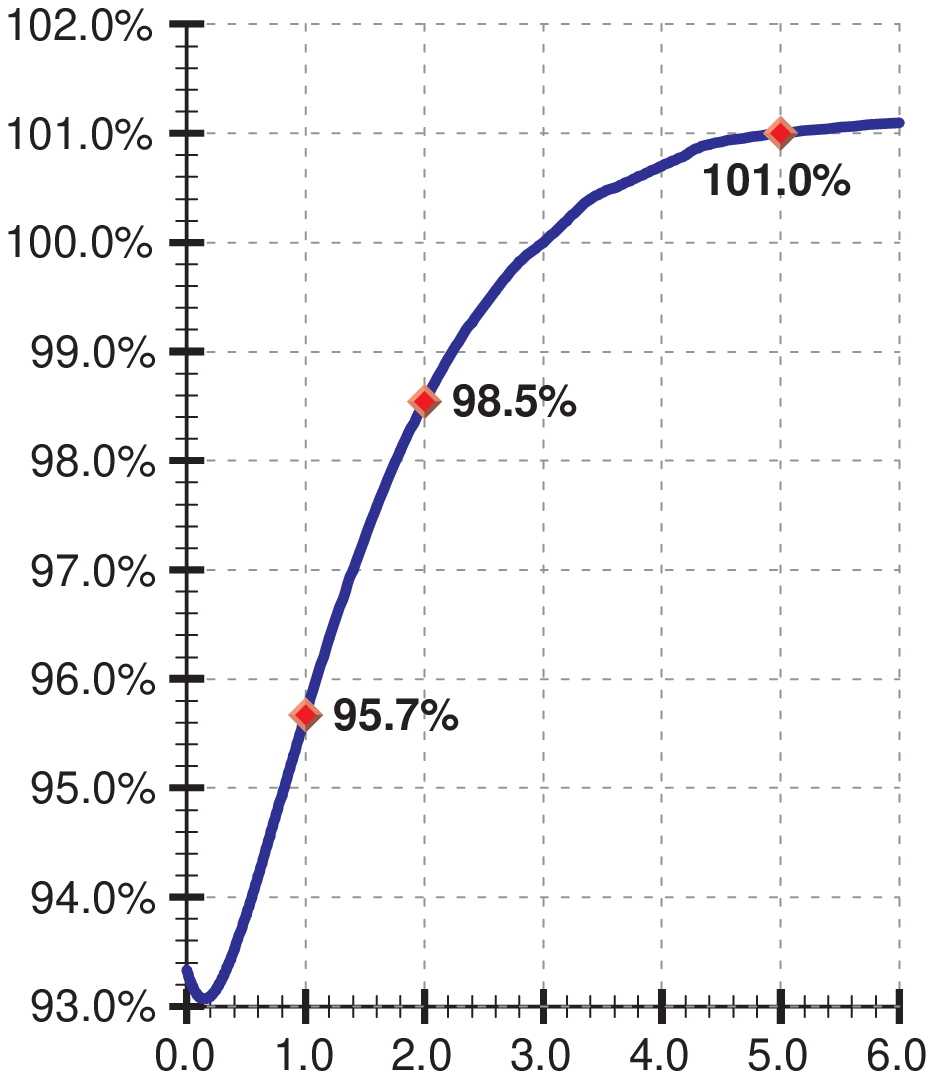, width=3.4in}
\put(-45,0){\Large\textsf{Depth,} $\ell$}
\put(-88,50){\rotatebox{90}{\Large$\mathbf{\mathsf{R_L(h)}}$}}
\caption{Ratio $R_L(h)$ between the integral \ic\ \I\ given by the \DA\ and full integral \ic\ \I\ for various depths, where the integration is taken over $4\pi$-steradians.}
\label{IntL}
\end{figure}

\begin{figure}
\epsfig{file=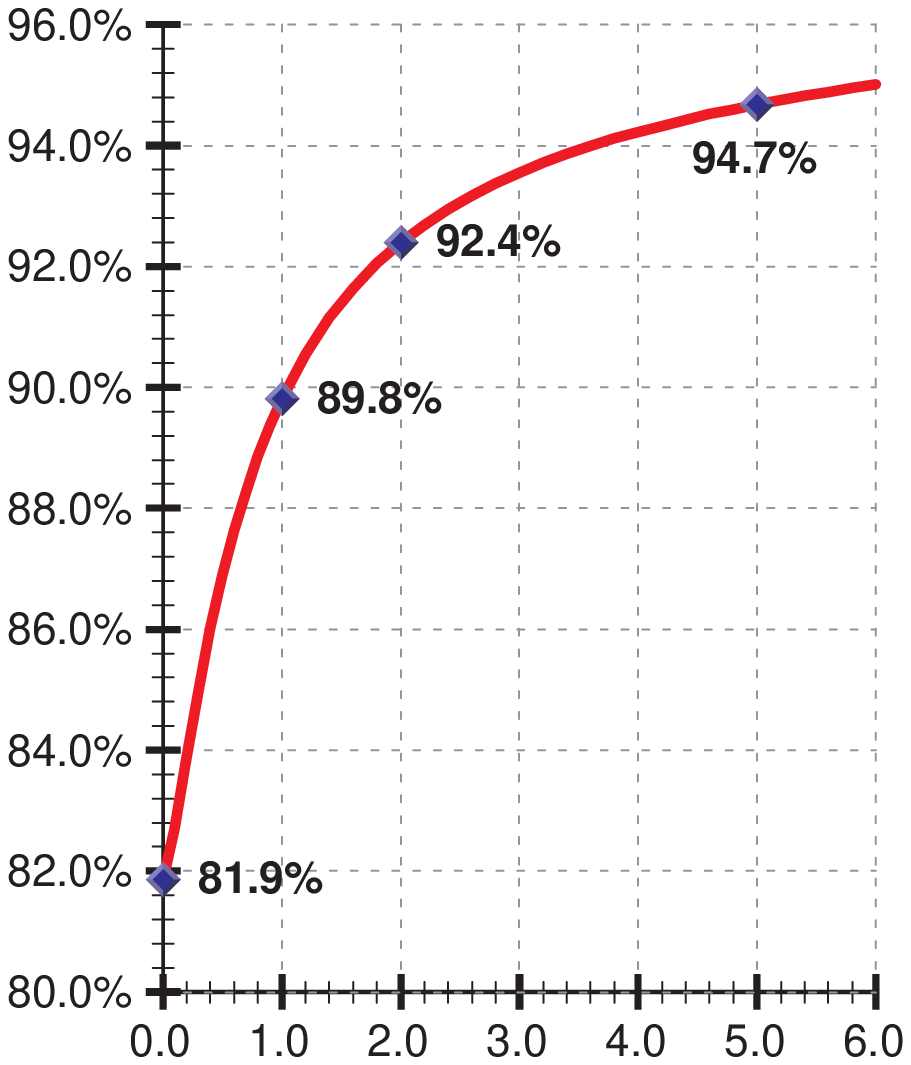, width=3.3in}
\put(-45,0){\Large\textsf{Depth,} $\ell$}
\put(-88,50){\rotatebox{90}{\Large$\mathbf{\mathsf{R_C(h)}}$}}
\caption{Ratio $R_C(h)$ between the integral \coh\ \I\ given by the \DA\ and full integral \coh\ \I\ for various depths, where the integration is taken over $1/k\el$.}
\label{IntC}
\end{figure}

\begin{figure}
\epsfig{file=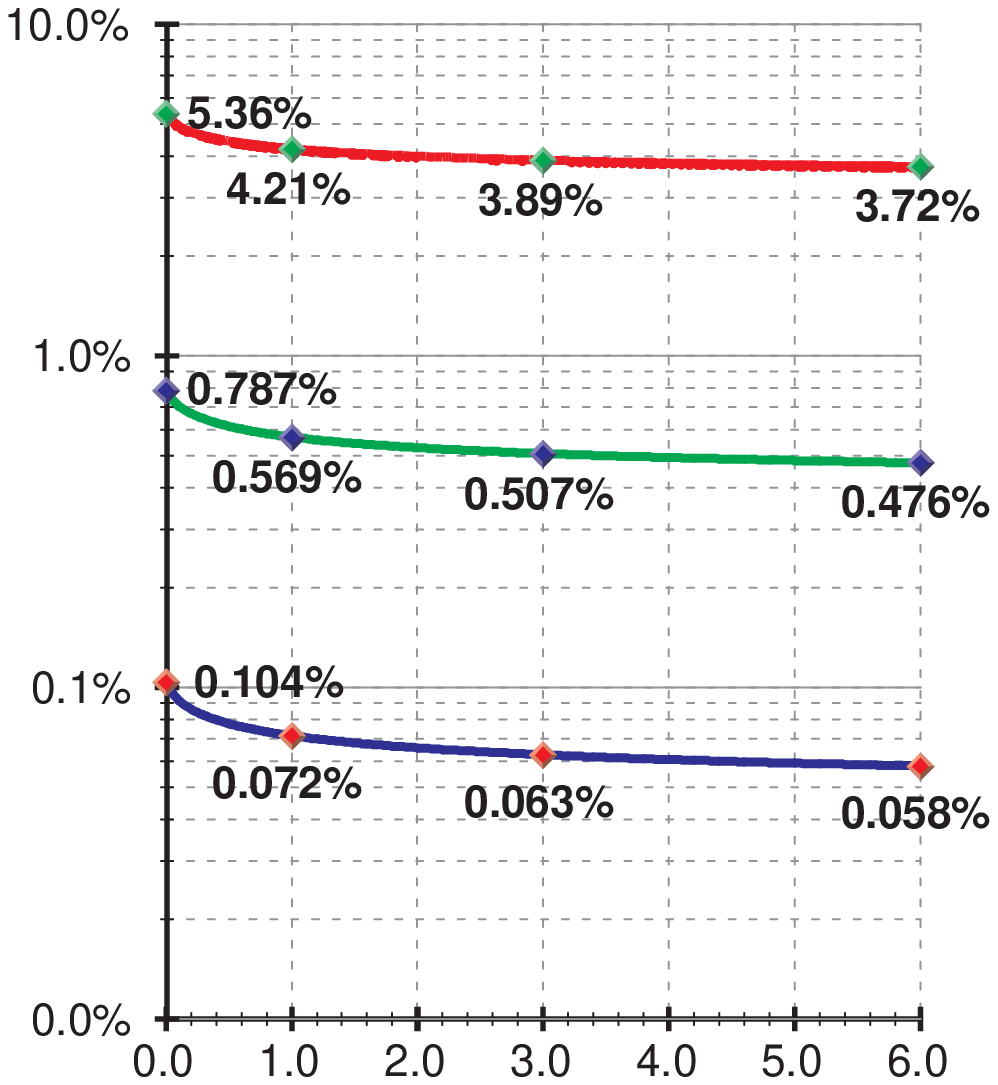, width=3.4in}
\put(-45,0){\Large\textsf{Depth,} $\ell$}
\put(-35,20){\textcolor{blue}{\Large$k\el=10^4$}}
\put(-35,45){\textcolor{green}{\Large$k\el=10^3$}}
\put(-35,70){\textcolor{red}{\Large$k\el=10^2$}}
\caption{Ratio between the integral \coh\ \I\ and integral \ic\ \I\ for various depths and various $k\el$.}
\end{figure}

As it was expected there is not any significant difference between the integral contribution of the first term, which corresponds to the \DA , and the second one in the \Eq{JLS}. The relative integral contribution of the \DT\ is given by the ratio \Eq{RL}, graphically represented on the Fig.~\ref{IntL}. As it is remarked above the lower order \dg\ are not correctly accounted by \DA. Very far from the surface, where unscattered \I\ is already vanished, the main contribution to the \I\ is due to the very high order \ldg. That's why in this case the \DA\ brings the exact result (see Fig.~\ref{IntL}).

The similar results are obtained for the \coh\ case (ratio \Eq{RC}) when the integration is taken over narrow cone around the \BSC\ peak ($\alpha_0=1/k\el$). The numerical calculations represented on Fig.~\ref{IntC} show that for the characteristic polar angle interval ($\Br{ 0,\,1/k\el}$) the \DA\ is permissible.

Unfortunately the {\bf integral value of the \coh\ \I\ is not concentrated only near the \BSC\ peak}. It is interesting that for the large polar angles the values of the \fu\ \Eq{JCS} are extremely low, and in spite of this {\bf the main contribution in the integral \coh\ \I\ comes from the integration outside the characteristic cone}. It is obviously from the numerical results (Fig.~\ref{IntC}), where the integration is taken over all directions ($\alpha_0\!=\!\pi$), that the contribution of the \DT\ in the \coh\ \I\ (ratio \Eq{RC}) is negligible. Moreover, as larger is the parameter $k\el$ as lower is the integral contribution of the \DT\ to the whole integral \I\ coming from the \cdg.

Finally on Fig.~\ref{IntC} is represented the ratio, multiplied by $k\el$, between the integral magnitudes of the \coh\ and \ic\ scattered \I. It is naturally, that the integral \coh\ \I\ is proportional to the perturbation theory parameter $1/k\el$, moreover this proportionality becomes linear for sufficiently small $1/k\el$. But the ratio, shown on Fig.~\ref{Int3}, also depends on $k\el$, so the above mentioned relationship is not fulfilled in the \DA. An important conclusion, we can derive from Fig.~\ref{IntC}, is that the \CSC\ is an effective surface phenomenon.

\section{Conclusion.}
The main result of the present paper are the exact \ex s for the \ic\ \Eq{JL} and \coh\ \Eq{JC} \I\ of a scalar wave scattered by semi-infinite disordered media in the general case of moving scatterers. The intensities are found as functions of the observation point and wave vector direction for all values of the solid angle ($4\pi-$ geometry). This allows us to analyze critically all the \app s done previously and to evaluate their deviation from the exact result.

In the case of \MSC\ in disordered media the far most popular \app\ is the \DA. However although in some papers \cite{Gb,AWM,McKSJ} it was mentioned that the \DA\ does not give correct results even for the simplest case of \ic\ \I, which corresponds to the summation of all ladder graphs in the mass operator, it does not exist analysis of the deviation of the exact results from these obtained in the \DA. We performed this kind of analysis in several important cases.

For the simplest case of statistical homogeneous and isotropic medium of static point-like ideal scatterers we found that the systematic error due to the \DA\ for the \ic\ \I\ on the surface and for various directions is about 6-7\,\%. We obtained that the error decreases in depth of the half space and at infinity becomes zero.

For the case of \coh\ \sc\ it is well known that the diffusion \app\ gives acceptable results for small angles around $\pi$ inside the region $\pi\+1/k\el$. In this paper we demonstrated quantitatively the deviation from the exact result. Even in backward direction on the surface of the medium the \DA\ does not reproduce the exact result but gives the same error (7\,\%) as in the \ic\ case. Away from the backward direction the relative contribution of the diffusion \app\ term of the \coh\ \I\ dramatically decreases. Our \ex\ \Eq{JC} gives the possibility to calculate the \coh\ \I\ for any value of the \sc\ angle including in forward direction. For this particular case the \DA\ is an order of magnitude smaller than the exact result.

In the case when the multiple \CBSC\ is also taken into account the \DA\ for the \I\ breaks down for large deviations from backward direction. The usual explanation of this fact is the incorrect description in this case of the interference effects for \sc\ with low multiplicity order \cite{Gb,AWM}. In order to check this conjecture we had to obtain explicit and exact \ex s for the intensities with fixed multiplicity order ( "partial" intensities ). A natural starting point to do this are the derived in Chapters V and VI exact \ex s for the \ic\ and \coh\ part of the intensities. Then the problem is reduced to correct and consistent transition from the \DA\ of the exact propagator to the \DA\ of the propagators with given order of multiplicity of the \sc. The consistency is the fulfillment of the condition that with increasing the number of the partial intensities which we sum up the sum should converge to the total \I\ scattered at given angle. Until present work the only known results in this direction were the obtained in the \ex s for the partial \coh\ intensities in the case of static point scatterers. However they did not fulfil the above formulated consistency condition. The key to the solution of this problem is the understanding that {\bf the \DA\ considered as asymptotic case for $R\!\gg\!\el$ \cite{Dev} do not match with the spread opinion that it corresponds to an \app\ for the \FI\ $\LL(K,t)$, where for $K\!\ll\!1/\el$ \ $\arctg{K\el}/K\el\!\approx\!1\-(K\ell)^2/3$}. It is clear that the cut off in the variable $K$ is not a correct procedure, when representing $\LL(R,t)$ with its \FI\ $\LL(K,t)$. This leads to the wrong guess for the geometrical series base for $\LL(K,t)$ and therefore to wrong results for the partial intensities for a given \sc\ angle in \DA\ \cite{AWM}. The proposed by us procedure for the \DA\ of the partial intensities is consistent. For example the sum of the first 30 multiplicity orders for the multiple \CBSC\ \I\ give 60.1602\,\% for angle $\pi\-\alpha:=0$ and 99.9972\,\% for angle $\pi\-\alpha:=\lambda/2\pi\el$.

The comparison of the \coh\ intensities for 2, 3, 4, 5, 6 and 7 multiplicity orders of \sc\ calculated with the exact formula and our formula in \DA\ demonstrates an increase of the relative deviation $(J^C_n\-^dJ^C_n)/J^C_n$ with increasing the angle $\pi\-\alpha$. For a fixed value of the angle the relative deviation increases with increasing of the multiplicity order. In the same time because of the sharp decrease of the contribution of a given multiplicity order the deviation $(J^C_n\-^dJ^C_n)/J^C$ is largest in case of double \sc. Only in this sense one may say that the low multiplicity orders of \sc\ are cut for large values of $\pi\-\alpha$.

When we investigated the behaviour of the integral \ic\ \I , we found the same degree of validity of \DA\ as the differential \I. But it is more interesting for us, what is happened with the integral \coh\ \I. If the polar angle scale, for which the \DA\ is acceptable, is limited to $1/k\el$, the corresponding solid angle scale is $\leq\!\pi/k^2\el^2$. That's why the \DA\ gives a {\bf drastic deviation from the correct values of the integral \coh\ \I\ } (Fig.~\ref{}), compared to the deviation of the differential \coh\ \I\ for a fixed large angle. It is noteworthy that the integral \coh\ \I, calculated by the exact \ex, depends linearly on the perturbation parameter $1/k\el$ for $k\el\!\gg\!1$. This dependence is not satisfied in the \DA. When we set ourselves the problem for the energy losses due to \CSC, we should have in mind that, {\bf the main integral \I\ originates from the \sc\ directions far from the \BSC\ peak}. That's why if someone is interesting in such item he will have to integrate the exact \ex\ \Eq{JC} in $4\pi$-geometry.

We have to remark, that the represented results for \sc\ \I\ are obtained for Miln boundary problem, solved in \DA\ \cite{Dev}, where the Miln number is $0,7104$. However the exact value of the Miln number is a bit different, but in our opinion it could not essentially influence on the represented numerical results.

The way of obtaining the exact results for the case of half-space, allows an extension of the method to the case of flat or spherical slab. In the standard approach the authors \cite{Gb,AWM,McKSJ} found the explicit form of the \coh\ \pr\ in space variables as a \FI\ of the time-reversed \lpr\ in wave-vector \rep. After that they replaced it in the \coh\ \I\ \ex.  In this work we perform a second operation of time-reversal directly inside the \coh\ \I\ \ex. The last is expressed by the \lpr\ in space variables, which have the same sequence like this one in the \ic\ \I\ \ex\ and the \GF\ variables have the exchanged places.

{\bf Our formulas conserve their validity also for small \EDL}. That means, they can describe the case of fast nonrelativistic scatterer motions as well as strong absorption. These situations lead to a cut off the constructive interference for higher order \sc\ and suppressing of the \coh\ effects.

If somebody keep his attention and interest in the content of this paper, he could have a background for an improvement of the popular results for the \MSC\ of the polarized light or to think about how it could be taken into account the scatterer size in further investigations.

\section{Acknowledgments}
The author would like to thank  Prof. Matey Mateev, who introduced me to the problem and helped me with encouragement and numerous  fruitful discussions. Also I am grateful to Drs M. Kostov and J. Velev for scientific and technical support and to Ch. Palamarev, who help me with software.

\newpage
\widetext

\appendix
\section*{}
\subsection{Calculation of the $L_n$ in space variables.}
\def\Lght{15}
\def\COORDIN#1#2#3#4{
\unitlength=1.6mm \lth{1.0pt}
\begin{picture}(80,22)(#1,#2){
\put(#3,#4){%
\put(-9,-4){\makebox(0,0)[cc]{{\large b}}}
\put(0,2.5){\makebox(0,0)[cc]{{\large $\circlearrowleft$}}}
\put(4,2.5){\makebox(0,0)[cc]{$i\chi(t)$}}
\put(0,2.5){\circle*{0.7}}
}%
\put(-9,-4){\makebox(0,0)[cc]{{\large a}}}
\multiput(0,0)(#3,#4){2}{%
\put(-\Lght ,0.5){{\lth{0.4pt} \vector(1,0){30}}}
\put(\Lght ,-2){\makebox(0,0)[cc]{$Re$}}
\put(-0.7,-5){{\lth{0.2pt} \vector(0,1){20}}}
\put(3,15){\makebox(0,0)[cc]{$Im$}}
\put(1,1){\oval(20,20)[tr]}
\put(-1,1){\oval(20,20)[tl]}
\put(0,6.0){\oval(2.0,2.0)[b]}
\put(1,11){\vector(0,-1)5}
\put(0,6){\vector(0,1)9}
\put(-1,6){\vector(0,1)5}
\multiput(-11,1)(8,0)2{\vector(1,0){8}}
\put(5,1){\line(1,0){6}}
\put(2,6){\makebox(0,0)[cc]{$i$}}
\put(0,6){\circle*{0.5}}
\put(11,11){\makebox(0,0){$R_{\infty}$}}
\put(7,7){{\lth{0.2pt}\vector(1,1){2.5}}}
}%
}
\end{picture}
}
\begin{center}
\COORDIN{-17}{-5}{45}{0}
\end{center}

\def\Lim#1#2{\lim_{#1\rightarrow#2}}
\def\eif{\e^{i\varphi}}
\def\sf{\sin\varphi}
\def\cf{\cos\varphi}

\def\ATGue#1{
\frac{\e^{(-u{#1}i\ep)\fRl}\arctg(iu{#1}\ep)}{\GT-\Atg{iu{#1}\ep}} }
\def\ATGz#1#2{
\frac{\arctg(|z|\eif)\,\e^{{#1}\fRl |z|\sf}\e^{{#2}i\fRl|z|\cf}i\eif }
{\GT-\Atg{|z|\eif}}\,d\varphi}

\def\Raz#1#2{\int_{#1\ep}^{#2\ep}\e^{i\fRl z}\br{\frac{\arctg z}z}^{\!n}z\d{z}}
\def\ARC#1#2{\Lim{|z|}\infty\int_{#1}^{#2}\e^{i\fRl|z|\eif}
\,\Atg{|z|\eif}\,|z|^2i\e^{2i\varphi}\d\varphi }
\widetext
\Beq[A1]
\lefteqn{
\iLn=\;\Lim\ep{0_+}
\left\{
\ARC\pi{\arccos\frac{-\ep}{|z|}}\;+
\right.}\NN
& &+\Raz{+i\infty-}{i-}
+i\ep\!\int_{-\pi}^0\e^{-\fRl (1-i\ep\eif)}
\Br{\Atg{i+\ep\eif}}^n(i+\ep\eif)\eif\d\varphi\;+ \NN
& &\left.
+\Raz{i+}{i\infty+}+\ARC0{\arccos\frac\ep{|z|}}
\right\}
\Eeq


Only second and fourth terms give nonzero yield in the integral \Eq{A1}, so

\def\Rau#1#2{\frac{\e^{-\fRl(u#1i\ep)}}{(iu#2\ep)^{n-1}}\Br{\arctg(iu#2\ep)}^n }
\Beqn
\lefteqn{
\iLn=\;\Lim{\delta}{0_+}i\!\int_{1+\delta}^\infty \Lim\ep{0_+}
\BR{\Rau-+\!- \Rau+-}\du=
}\NN
& &= \INT\frac{\e^{-\fRl u}}{2^n u^{n-1}}\Br{\br{\LP+i\pi}^{\!n}
\!-\br{\LP-i\pi}^{\!n} }\du= i\!\PIterm{\,\e^{\fRl u}}
\Eeqn
Replacing this result in \Eq{LnI} we will have the following \ex s for
first 6 bare \ldg:

\Beq[A2]
L_2(R,t)&=&\frac{g^2(t)}{\Rl}\INT\E{R}\du=\frac{g^2(t)\exp\br{-\fRl}}{R^2\el^2} \NN
\Lpar3{{u}\LP\,}
\Lpar4{{(2u)^2}\br{3\LPa}}
\Lpar5{{2u^3}\LP\br{\LPa}}
\Lpar6{{16u^4}\br{5\LPb\-\LPc\+\pi^4}}
\Lpar7{{16u^5}\LP\br{3\LPb\-\LPc\+ 3\pi^4}}
\Eeq

\subsection{Calculation of the $\LL$ in space variables.}

\Beq
\lefteqn{
\InT\frac{\arctg\,p\:\e^{ip\fRl}\d{p}}{\GT-\frac{\arctg p}p}=\;\;
2\pi i\;{ {\rm Res}\frac{\arctg\,z\,\exp\br{-iz\fRl}}
{\GT-\Atg{z}} }_{\!|\,z=i\chi}\!\!\!+
}\NN
& &+\Lim\ep{0_+}\left\{ i\!\INT\,\Br{\ATGue+ -\ATGue- }\du +
\int_{-\pi}^0\frac{\arctg(i+\ep\eif)\,\e^{-\fRl(1+\ep\sf)}\,\e^{i\fRl\ep\cf}
\,i\ep\eif}{\GT-\Atg{i+\ep\eif}}\d\varphi - \right. \NN
& &-\left.{\Lim{|z|}\infty |z|\Br{\int_0^{\arccos\frac\ep{|z|}}
\ATGz-{} + \int_{-\arccos\frac\ep{|z|}}^0 \ATGz{}- } }\right\}
\Eeq

\subsection{$3D\:\delta$-function.}
\def\bRR{\sqrt{X_1^2+(\ZZ_1)^2}}
\Beq
\lefteqn{\DEks1=}\NN
& &=\DE\br{k\SIA+k_0\frac{X_1}\aRR}\,\DE\br{k_0\frac{Y_1}\aRR}
\,\DE\br{k\COA-k_0\frac{\ZZ_1}\aRR}= \NN
& &=\frac\bRR{k_0}\,\DE\br{k\SIA+\frac{k_0X_1}\bRR}\,\DE(Y_1)
\,\DE\br{k\COA-\frac{k_0(\ZZ_1)}\bRR}=\NN
\ \NN
& &=\frac{(\ZZ_1)^2}{k_0^2\,\cos^4\alpha}\,\left|\COA+\frac{\sin^2\alpha}{\MC}
\U{\ZZ_1}\right|^{-1}\TZ1\,\DE(k-k_0)\,
\DE\br{X_1+\SIA\frac{|\ZZ_1|}\MC}\,\DE(Y_1)=\NN
\ \NN
& &=\left\{
\begin{array}{ll}
\frac{\ZZ_1}{k_0^2\,\MC^3}\,\DE(k-k_0)\,\DE(X_1+\tga(\ZZ_1))\,
\DE(Y_1)\,\TZ1&;\;\;\alpha\neq\pi/2\\
\frac{X_1^2}{k_0^2}\,\DE\br{k+k_0\U{X_1}}\,\DE(Y_1)\DE(\ZZ_1)=
\frac{X_1^2}{k_0^2}\,\DE(k-k_0)\,\DE(Y_1)\,\DE(\ZZ_1)\,\Theta\br{-X_1}&;\;\;\alpha=\pi/2
\end{array}
\right.
\Eeq

\subsection{Calculation of single \sc\ \I.}
\beq
L_1(\Rab;\,k_0\Sv_1,k_0\Sv_2;\,t)=
\DE(\RaRb)\Fi{k_0\SS12,t}\approx\frac{4\pi}\el\,\DE(\RaRb)\,g(t)
\eeq

\beq[GL1]
\Gamma_1\Rkt:=\int\!\exp(-i\kv\.\rv)\,\Gamma_{L_1}\Rrt\dr
=\frac{2\pi^2}\el\,g(t)\,\eot \int\!\DE(\kv -k_0\Sv_{RR_1})\,\DDR1\,
|E(\Rv_1)|^2\dR_1
\eeq

\beq
J_1(Z,,\alpha,t)=\frac{g(t)\NORM}{2\el\,\MC}\int_0^{+\infty}
\TZ1\,\EZ1\,\e^{-\frac{Z_1}\el}\dZ_1
\eeq

\beq
J_1\hat 2\NORM\,g(t)\,\frac{\e^{-h}-\TECA}{1-\COA}
\eeq

\subsection{Albedo}
\beq[HLO]
H^C(0,\alpha,u)=\frac1{\ZNA}\!
\br{\frac{\COA }{\COA\- 1}+\frac{1\-\e^{-M\,\xi}}{\xi}};\;\;\;\;\frac\pi2\leq\alpha\leq\pi
\eeq

\subsection{Calculation of partial intensities.}

\def\uu{u}
\def\pP{\frac{(\uu\+1)^{m-k}-(\uu\-1)^{m-k}}2}
\Beq[p20]
\lefteqn{
J_{n+1}^{L^{\rm diff}}(h, 0, t)= 2\NORM\gt
\left\{\; \br{\frac{3\+\Xt1}{2\+\Xt1}}^{\!n}h\,\e^{-h}+
\hspace{60pt}u:=\sqrt{3\+2\Xt1}
\right.}\NN
& &+\,(3\+\Xt1)^n\!\sum_{k=1}^n{2n\-k\-1 \choose n\-1}(2\uu)^{k-2n}
\sum_{m=0}^{k-1}\frac1{m!}
\left[(k\-m)(\uu\-1)^{m-k-1}\br{h^m\e^{-\uu h}-\theta(-m)\e^{-h}}-\right.\NN
& &\left.\left.-\,\pP\br{ (h\+2\Mas)^m\exp\br{-\uu(h\+2\Mas)}-
(2\Mas)^m\exp\br{-2\uu\Mas\-h} }\right]\right\}
\Eeq
\ \\
\Beq[p21]
\lefteqn{
J_{n+1}^{L^{\rm diff}}(h, \frac\pi2, t)= 2\NORM\gt
\left\{\; \br{\frac{3\+\Xt1}{2\+\Xt1}}^{\!n}\e^{-h}\,+
\right.}\NN
& &\left.+\,(3\+\Xt1)^n\!\sum_{k=1}^n{2n\-k\-1 \choose n\-1}(2\uu)^{k-2n}
\sum_{m=0}^{k-1}\frac1{m!}\!\left[h^m\e^{-\uu h}(\uu\-1)^{m-k}
+(h\+2\Mas)^m\exp\br{-\uu(h\+2\Mas)}(\uu\+1)^{m-k}
\right]\right\}
\Eeq
\ \\
\Beq[p22]
\lefteqn{
J_{n+1}^{L^{\rm diff}}(h, \pi, t)= 2\NORM\gt
\left\{\; \br{\frac{3\+\Xt1}{2\+\Xt1}}^{\!n}\frac{\e^{-h}}2
+\,(3\+\Xt1)^n\!\sum_{k=1}^n{2n\-k\-1 \choose n\-1}(2\uu)^{k-2n}\times
\right.}\NN
& &\left.\times\sum_{m=0}^{k-1}\frac1{m!}\!\left[h^m\e^{-\uu h}\pP-
(h\+2\Mas)^m\exp\br{-\uu(h\+2\Mas)}(k\-m)(\uu\+1)^{m-k-1}
\right]\right\}
\Eeq
\end{document}